\documentclass[twoside,twocolumn,9pt]{article}
\usepackage{extsizes}
\usepackage[super,sort&compress,comma]{natbib} 
\usepackage[version=3]{mhchem}
\usepackage[left=1.5cm, right=1.5cm, top=1.785cm, bottom=2.0cm]{geometry}
\usepackage{balance}
\usepackage{times,mathptmx}
\usepackage{sectsty}
\usepackage{graphicx} 
\usepackage{lastpage}
\usepackage[format=plain,justification=justified,singlelinecheck=false,font={stretch=1.125,small,sf},labelfont=bf,labelsep=space]{caption}
\usepackage{float}
\usepackage{fancyhdr}
\usepackage{fnpos}
\usepackage[english]{babel}
\usepackage{array}
\usepackage{droidsans}
\usepackage{charter}
\usepackage[T1]{fontenc}
\usepackage[usenames,dvipsnames]{xcolor}

\usepackage{setspace}
\usepackage[compact]{titlesec}
\usepackage{hyperref}

\usepackage{graphics}
\usepackage{mathptmx}
\usepackage[squaren,Gray]{SIunits}
\usepackage{amsfonts}
\usepackage{amssymb}
\usepackage[squaren,Gray]{SIunits}

\usepackage{epstopdf}

\definecolor{cream}{RGB}{222,217,201}

\begin{document}

\pagestyle{fancy}
\thispagestyle{plain}
\fancypagestyle{plain}{

}

\makeFNbottom
\makeatletter
\renewcommand\LARGE{\@setfontsize\LARGE{15pt}{17}}
\renewcommand\Large{\@setfontsize\Large{12pt}{14}}
\renewcommand\large{\@setfontsize\large{10pt}{12}}
\renewcommand\footnotesize{\@setfontsize\footnotesize{7pt}{10}}
\makeatother

\renewcommand{\thefootnote}{\fnsymbol{footnote}}
\renewcommand\footnoterule{\vspace*{1pt}%
\color{cream}\hrule width 3.5in height 0.4pt \color{black}\vspace*{5pt}} 
\setcounter{secnumdepth}{5}

\makeatletter 
\renewcommand\@biblabel[1]{#1}            
\renewcommand\@makefntext[1]%
{\noindent\makebox[0pt][r]{\@thefnmark\,}#1}
\makeatother 
\renewcommand{\figurename}{\small{Fig.}~}
\sectionfont{\sffamily\Large}
\subsectionfont{\normalsize}
\subsubsectionfont{\bf}
\setstretch{1.125} 
\setlength{\skip\footins}{0.8cm}
\setlength{\footnotesep}{0.25cm}
\setlength{\jot}{10pt}
\titlespacing*{\section}{0pt}{4pt}{4pt}
\titlespacing*{\subsection}{0pt}{15pt}{1pt}

\fancyfoot{}
\fancyfoot[RO]{\footnotesize{\sffamily{1--\pageref{LastPage} ~\textbar  \hspace{2pt}\thepage}}}
\fancyfoot[LE]{\footnotesize{\sffamily{\thepage~\textbar\hspace{3.45cm} 1--\pageref{LastPage}}}}
\fancyhead{}
\renewcommand{\headrulewidth}{0pt} 
\renewcommand{\footrulewidth}{0pt}
\setlength{\arrayrulewidth}{1pt}
\setlength{\columnsep}{6.5mm}
\setlength\bibsep{1pt}

\makeatletter 
\newlength{\figrulesep} 
\setlength{\figrulesep}{0.5\textfloatsep} 

\newcommand{\topfigrule}{\vspace*{-1pt}%
\noindent{\color{cream}\rule[-\figrulesep]{\columnwidth}{1.5pt}} }

\newcommand{\botfigrule}{\vspace*{-2pt}%
\noindent{\color{cream}\rule[\figrulesep]{\columnwidth}{1.5pt}} }

\newcommand{\dblfigrule}{\vspace*{-1pt}%
\noindent{\color{cream}\rule[-\figrulesep]{\textwidth}{1.5pt}} }

\makeatother

\twocolumn[
  \begin{@twocolumnfalse}
\vspace{3cm}
\sffamily
\begin{tabular}{m{4.5cm} p{13.5cm} }

\includegraphics{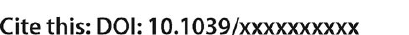} & \noindent\LARGE{\textbf{Distribution of active forces in the cell cortex}} \\
\vspace{0.3cm} & \vspace{0.3cm} \\

 & \noindent\large{P. Bohec$^{\ast}$, J. Tailleur, F. van Wijland, A. Richert and F. Gallet} \\

\includegraphics{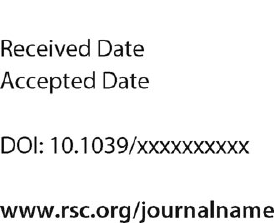} & \noindent\normalsize{In this work, we study in detail the distribution of stochastic forces generated by the molecular motors activity, in the actin cortex of pre-muscular cells. By combining active and passive rheology experiments, performed on the same micro-bead bound to the actin network through membrane adhesive receptors, we measure the auto-correlation function $C_{ff}(\tau)$ of the average force pulling on the bead. Like for any out-of-equilibrium system, the force distribution differs from the thermodynamical equilibrium one, especially at long time scale $\tau \gtrsim 1 \sec $ where the bead motion becomes partially directed. Thus the fluctuation-dissipation theorem does not apply and one can measure the distance from equilibrium through its violation. This work focuses on the influence of various parameters (ligand density, temperature, ATP depletion, molecular motors activity) on the force distribution. In particular, it is shown that the amplitude of active forces increases when the bead is more tighly attached to the cortex: this is interpreted through a model which takes into account the number of bonds between the bead and the cytoskeleton and the viscoelastic properties of the medium. It also increases with temperature, consistently with a description of the cell metabolism in terms of thermally activated reactions. Last, but not least, ATP depletion in the cell, or partial inhibitition of the actomyosin activity, leads to a decrease of the amplitude of the force distribution. Altogether, we propose a consistent and quantitative description for the motion of a micrometric probe interacting with the actin network, and for the amplitude of the stochastic forces generated by molecular motors in the cortex surrounding this probe.} \\

\end{tabular}

 \end{@twocolumnfalse} \vspace{0.6cm}

  ]

\renewcommand*\rmdefault{bch}\normalfont\upshape
\rmfamily
\section*{}
\vspace{-1cm}


\footnotetext{\textit{ Mati\`ere et Syst\`emes Complexes, UMR 7057 associ\'ee au CNRS et \`a l'Universit\'e Paris-Diderot, 10 rue Alice Domon et L\'eonie Duquet, F-75013 Paris. E-mail: francois.gallet@univ-paris-diderot.fr}}



\footnotetext{$^{\ast}$\textit{Present adress: London Centre for Nanotechnology, University College London, London WC1E6BT, UK.}}


\section{Introduction}
Biological activity of living cells is characterized by permanent exchanges of matter and energy with the environment, while biochemical reactions provide the power supply necessary to the metabolism. This activity allows for cell motion and deformation required for function or survival, and cargo transport within the dense microenvironment of the cell. These mechanical effects are directly related to the dynamics of polymerization/depolymerization of the cytoskeleton and to the activity of molecular motors, which convert the chemical energy stored in ATP into mechanical energy. Living cells are thus intrinsically out-of-equilibrium  systems, for which the thermodynamic temperature is no longer the sole pertinent parameter to characterize stochastic events. Consequently, the fluctuation-dissipation theorem does not apply and its violation can be used to quantify the distance from equilibrium: the motion of a freely diffusing probe embedded inside a living system cannot simply be  related to thermal fluctuations, but it is expected to be controlled by the active forces generated during cytoskeleton remodelling and cell trafficking. 

Evidences for such out-of-equilibrium behaviour have been described in recent works, either in living cells \cite{Lau2003,Hoffman2006,bursac_cytoskeleton_2007,Wilhelm2008,gallet2009,Mizuno2009,robert_vivo_2010,Turlier_2016} or in reconstituted active gels \cite{Mizuno2007,Brangwynne2008}. In these works, a combination of active and passive microrheology has generally been used to characterize the mechanical and rheological properties of the system. Usually, in active microrheology, the complex viscoelastic modulus $G(\omega) = G'(\omega)+iG"(\omega)$ of the medium is determined by applying an oscillating force at frequency $\omega$ to a micrometric probe embedded into or attached to the active network. Alternatively, the response function $J(t)$ characterizes the probe displacement in response to a step of force of fixed amplitude. In living cells, active rheology experiments are usually performed either at the cell cortex, by using beads specifically bound to transmembrane receptors and manipulated with optical tweezers or magnetic tweezers/twisters \cite{Fabry2001,Fabry2003,Balland2005}, or in the bulk cellular body, by using magnetic beads or magnetic endosomes embedded in the cytoplasm \cite{Bausch1999,Wilhelm2003}.
Most of these works report a weak power law dependence of the complex viscoelastic modulus with frequency, or of the creep function with time, and there is now a consensus on such behavior in the community \cite{Hoffman2006}.

On the other hand, passive microrheology allows one to retrieve the mean square displacement (MSD) of a freely-moving particle as a function of time. Different regions of the cell can be probed, using endogenous probes \cite{yamada_2000,aubertin_2017, ahmed2018active,fodor2016nonequilibrium}, or submicron-sized particles embedded within the cell body \cite{Tseng2002,Lau2003,Caspi2002,Gal2010,fodor2015activity}, or particles specifically attached to the plasma membrane \cite{bursac_cytoskeletal_2005,bursac_cytoskeleton_2007,Raupach2007,Massiera2007,bohec2013probing}, or vesicles moving in the cortical region \cite{Huet2006}. In most cases, the mean square displacement of the particle exhibits two regimes: a subdiffusive one at short time scales (typically $\tau<1s$), when the particle is confined by the viscoelastic network, and a superdiffusive one at longer time scales, corresponding to a partially directed motion of the particle driven away by molecular motors through the network. At least in this regime, the MSD reflects non-thermal motion and does not obey the fluctuation-dissipation theorem.

As shown in \cite{Lau2003,gallet2009}, the autocorrelation function $C_{ff}(\tau)=\langle f_{r}(t)f_{r}(t+\tau)\rangle$ of the stochastic force $f_{r}(t)$ exerted on a probe can be retrieved by combining the results of active and passive microrheology. At thermal equilibrium, one has $C_{ff}^{eq}(\tau) = \Gamma(\tau)k_B T $, in which $\Gamma(\tau)$ represents the viscoelastic response kernel of the $\{$bead+cortex$\}$ system.



Actually, in living systems both thermal fluctuations $\xi(t)$ and forces generated by molecular motors activity $f_{a}(t)$ contribute to the random force $f_{r}(t)$. Previous works have shown that, at least for long time scales, the contribution of active forces $f_{a}(t)$ dominates, and that the autocorrelation function $C_{ff}(\tau)$ may lie several orders of magnitude above the thermal one. The ratio $\frac{C_{ff}(\tau)}{C_{ff}^{eq}(\tau)}  = \frac{T_{\text eff}}{T}$ may be used to define an effective, time dependent, energy scale $ k_B T_{\text eff}$ for the system, where the so-called effective temperature $T_{\text eff}$ may be 100-1000 times larger than the thermodynamical temperature $T$. Besides, evidence for a departure from thermal equilibium at short time scales is still controversial \cite{Mizuno2007,gallet2009}.

In this paper, we address several issues concerning the active
(non-thermal) distribution of forces exerted on a micron-sized probe
attached to the actin network in a living cell. These probes are beads
specifically bound to the actin cortex of premuscular cells, through
RGD-integrin recognition at the plasma membrane. For each bead, we
record successively the spontaneous fluctuations of its position at
the membrane, and its displacement in response to a step force applied
by an optical tweezers. The combination of both experiments leads to
the autocorrelation function $C_{ff}(\tau)$ of all stochastic forces
exerted on the bead. We quantify the influence of various parameters
(ligand density, temperature, ATP depletion, molecular motors
activity), and discuss in detail their consequences on this force
distribution.

First, by varying the density of RGD ligand on the bead, we quantify how the distribution of forces exerted on the bead is affected by the degree of attachment of the probe to the actin cortex. Our results demonstrate that, as the link to the actin network is strengthened, the amplitude of short timescale fluctuations decreases, while the motion at large timescale becomes more directed.

Secondly, by comparing our data at two different temperature ($25\celsius$ and $37\celsius$), we find that the contribution of active forces increases with temperature. The mechanical dissipation in the system, computed according to \cite{Harada_Sasa}, is consistent with a description of the cell metabolism in terms of thermally activated biochemical reactions.

And, last but not least, we show that ATP depletion in the cell, or reduced mechanical activity consequent to addition of a myosin II inhibitor (blebbistatin), leads to a decrease of the active part of the force distribution. 

In rationalize our findings by putting forward a model which describes in detail the mechanical connexion of the probe to the surrounding network. This model takes into account the number of links between the probe and the actin network, and predicts the active and passive responses of the probe. It leads to a qualitative agreement with the experimental results obtained by varying the ligand density. Altogether, we are able to give a consistent description of the motion of a micrometric probe interacting with this network, and to relate this motion to the distribution of forces generated by biological activity in the cell cortical network.

\section{Experimental set-up and methods}
\label{sec:methods}
\subsection{Cell culture and experimental chamber}
The muscular cell line C2C12 (kindly provided by M. Lambert and R.M. M\`ege, INSERM U440, Institut du Fer \`a Moulin, Paris) was maintained in a stock callus culture in Dulbecco's Modified Essential Medium (DMEM) with $2 \milli \mathrm{M}$ glutamine, $10\%$ fetal bovine serum and antibiotics (penicillin=$100~\mathrm{U} \per \milli \liter$, streptomycin=$50~\milli \gram \per \milli \liter$). Cells were grown at $37\celsius$ in a humidified $5\%$ CO$_2$ atmosphere. Just before confluent state, cells were subcultivated in a volume of $5 \milli \liter$ of fresh medium. The day before experiments, cells were detached using a buffer solution (PBS) with $0.05\%$ trypsin and $0.02\%$ EDTA and were plated on $22\textrm{x} 22~\milli \meter $ glass coverslips coated with $50~\micro \gram \per \milli \liter$ fibronectin. RGD-coated beads were then incubated on cells at $37\celsius$ for $30~\min$ before manipulation (3-4 beads per cell). After incubation, unbound beads were washed away with DMEM and each glass coverslip was sealed on a glass slide with an adhesive spacer (Gene Frame\textregistered, $1.5\textrm{x} 1.6~\centi \meter $). This makes an experimental chamber volume of $\ 65~\micro \liter $ filled with DMEM. The chamber was then attached to a piezoelectric stage in thermalisation box (The Box\textregistered, Life Imaging Services, Switzerland). For the experiments requiring a change of medium (section~\ref{MyosinIIinhibition}), cells were plated in a cover-glass bottom dish (hiwaki).

\subsection{Bead coating}
\label{Bead coating}
Carboxylated silica beads ($1.56~\micro\meter$ diameter, Bangs Laboratories, Fishers, IN) were coated with a polypeptide containing the Arg-Gly-Asp (RGD) sequence (PepTide 2000, Telios Pharmaceuticals, San Diego, CA), according to the manufacturer procedure. It ensured a specific binding to integrin receptors. We achieved different coating densities by changing the RGD concentration in solution during the incubation with beads. \textbf{RGD 1/10} corresponds to $40~\micro\gram \mathrm{ \ of \ peptide}\per\milli\gram$ of bead, \textbf{\textcolor{blue}{RGD 1/20}} to $20~\micro\gram \mathrm{ \ of \ peptide}\per\milli\gram$ of bead and \textbf{\textcolor{red}{RGD 1/40}} to $10~\micro\gram \mathrm{ \ of \ peptide}\per\milli\gram$ of bead. We assume that the adsorbed ligand density on the bead's surface increases with the ligand concentration in solution, allowing us to modulate the degree of attachment between the bead and the cell cortex. 

\subsection{Cell treatments}
\label{Cell treatments}
For ATP depletion, cells were incubated for $30~\minute$ in a solution of deoxyglucose ($6~\milli \mathrm{M}$) and NaN$_3$ ($10~\milli \mathrm{M}$). Then, cells were washed with a medium without serum. The experimental chamber was sealed without serum. To inhibit actin polymerization, cells were incubated for $30~\minute$ in a solution of latrunculin A ($1~\micro \mathrm{M}$). The experimental chamber was sealed with medium and latrunculin A. To inhibit Myosin II activity, cells were incubated for $30~\minute$ in a solution of blebbistatin ($50~\micro \mathrm{M}$). Then, the experiments were performed in a cover-glass bottom dish filled with medium and blebbistatin.
 
\subsection{Passive microrheology}
\label{Passive microrheology}
\begin{figure}[htb!]
  \begin{center}
  	\includegraphics[width=\columnwidth]{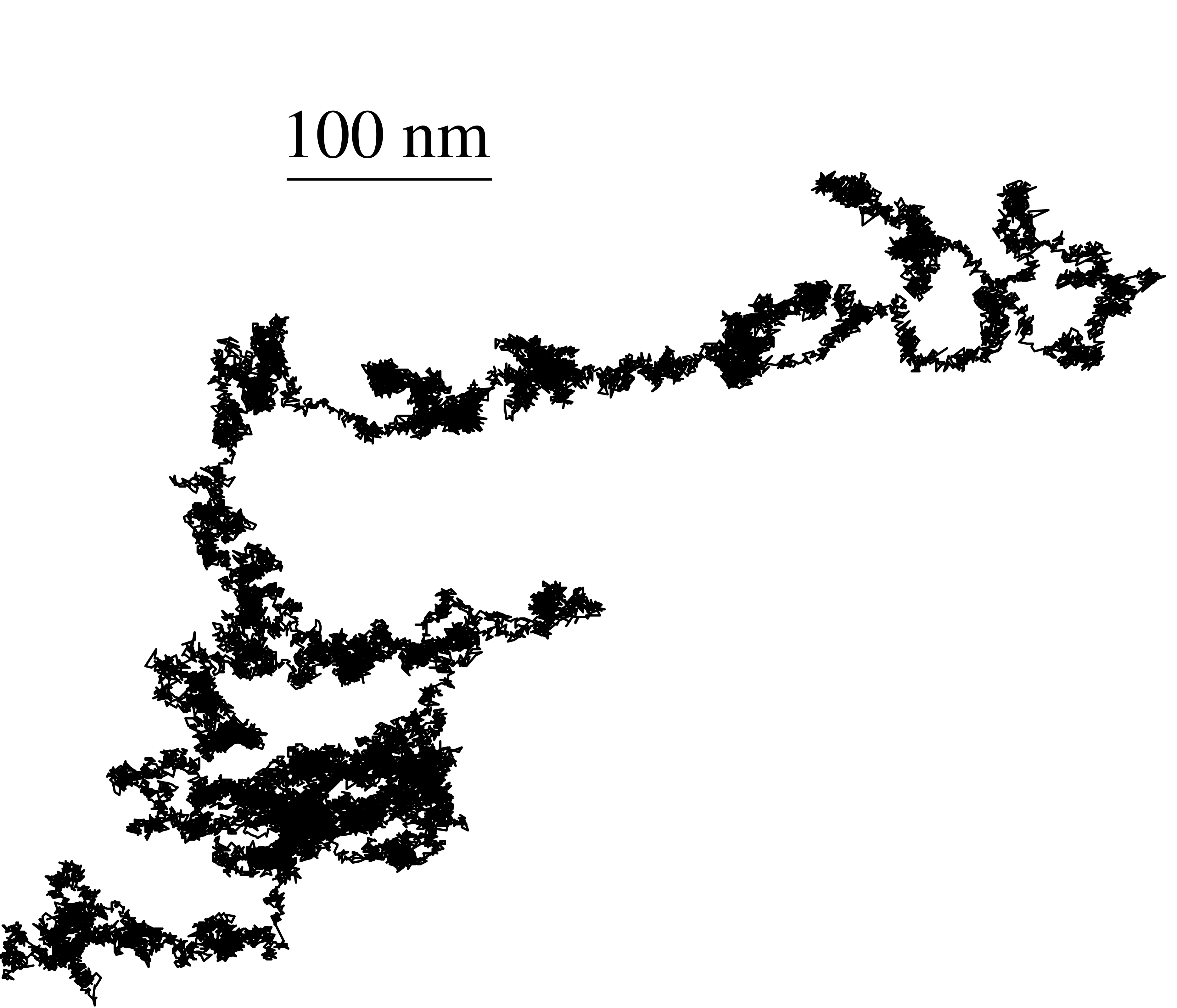}
    \caption{Example of recorded trajectory during the spontaneous motion of a bead bound to cortical actin \textit{via} integrins (time lapse $8~\minute~30~\second$).}
    \label{fig:trajectoire_bille}
  \end{center}
\end{figure}

To probe the passive microrheology behaviour, we record the bead's spontaneous diffusive motion along the membrane, and then calculate its mean square displacement. From the movie taken with a fast camera (FASTCAM-ultima 1024), the bead's positions as a function of time $\left\lbrace x_i,y_i\right\rbrace$ is extracted with a home-made particle tracking algorithm. A typical example of the spontaneous trajectory of a bead attached to cortical actin is shown in figure \ref{fig:trajectoire_bille}. 

For each trajectory, the mean square displacement (MSD) is calculated using:
\begin{eqnarray*}
\Delta x^2 (\tau)& = &\langle (x(t+\tau) - x(t))^2 \rangle_t \\			 
				 & = &\dfrac{1}{N-n} \sum_{i=1}^{N-n} (x_{i+n} - x_{i})^2.				 
\end{eqnarray*}

where $\tau = n\Delta t$ is the time lag, $\Delta t$ is the sampling time interval, and $N = 40000 $ is the total number of frames.
In these experiments, the sampling frequency  is either 125$\hertz$ ($N \Delta t = 8~\minute~30~\second$) or 250$\hertz$ ($N \Delta t = 4~\minute~15~\second$). To take into account the residual mechanical drift of the microscope stage, we measure during each run the displacement of a fixed bead attached to the microscope slide, and we subtract this drift from the recorded motion of the unperturbed beads. The noise level measured with the MSD amplitude of a fixed bead is $2.10^{-6}~\micro \meter^{2}$. Then, the MSD ensemble average over several realizations and the associated standard error (SE) are computed. Because of the log-normal distribution of MSD amplitudes, we calculate their geometrical average, as usually done in the literature \cite{aubertin_2017,Tseng2002,Lau2003,Caspi2002,Gal2010,bursac_cytoskeletal_2005,bursac_cytoskeleton_2007,Raupach2007,Massiera2007,Huet2006}.

\subsection{Active microrheology}
\label{Active microrheology}
Active microrheology makes it possible to probe the viscoelastic response of the bead's environment. Here the bead is manipulated with optical tweezers. The optical trap is created by an infrared Ytterbium fiber laser beam (IPG photonics, YLM-10-1050, 10W, TEM$_{00}$) focused through the oil immersion objective (Olympus UPlanFL 100x oil, NA = 1.30) of an inverted microscope (Olympus IX71). To first order, the trapping force $f_{\mathrm{ext}}$ exerted on a dielectric silica spherical bead is proportional to the bead-trap distance $x_{\mathrm{f}}$, and to the laser beam power. The trap stiffness $k$ can be accurately calibrated, following the procedure described in~\cite{Balland2005}.

\begin{figure}[h!]
  \begin{center}
  	\includegraphics[width=\columnwidth]{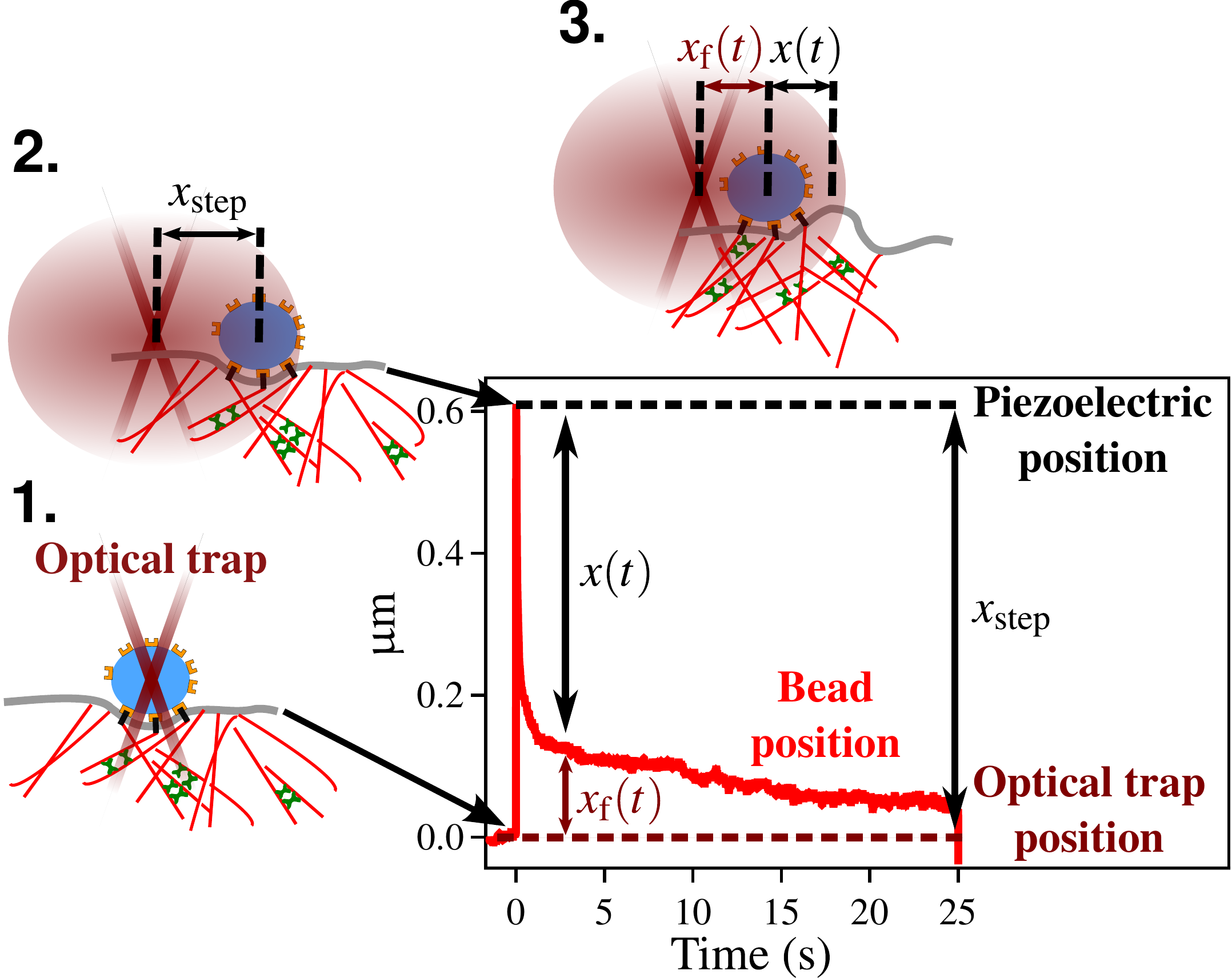}
    \caption{Typical sequence of active microrheology measurement}
    \label{fig:schema_fluage}
  \end{center}
\end{figure}

In our set-up, the position of the optical trap is fixed and a piezoelectric stage (Physik Instrumente PZ D-050) is used to move the sample with respect to the trap. A typical sequence of active microrheology measurement is drawn in figure~\ref{fig:schema_fluage}: at $t<0$, the bead is in equilibrium at the center of the trap (1). At $t=0^+$ (2), the piezoelectric stage performs an instantaneous displacement of the sample $x_{\mathrm{step}}=0.6 ~\micro \meter$, and a proportional step of force is applied to the bead. Then the bead relaxes towards the trap center (3), deforming its cellular environment. At  time $t$, the trapping force is $f_{\mathrm{ext}} = k x_\mathrm{f}(t)$ and the cell local deformation is $x(t)=x_{\mathrm{step}}-x_\mathrm{f}(t)$. In the following, the relationship between $x(t)$ and $f_{\mathrm{ext}}(t)$ is used to define the cell active response function $J(t)$ (see \ref{Formalism}).

\subsection{Formalism}
\label{Formalism}
We assume that the probe attached to the viscoelastic medium is
submitted to an effective friction force, characterized by a delayed
memory kernel $\Gamma(t)$, considered as isotropic for simplicity. As
we show in section \ref{Modelling the origin of the bead's active
  motion}, this can be derived from a microscopic model of the
connection between the bead and the cytoskeletton. Moreover, due to
the active properties of living matter, the bead experiences both
thermal and non-equilibrium biological stochastic forces
$\vec{f}_r(t)$ $=\vec{\xi}(t)+\vec{f}_{\mathrm{a}}(t)$. Neglecting the
bead's inertia, its dynamics is simply described by a
generalized Langevin equation:

\begin{equation}
\int_{-\infty}^{t}\Gamma\left(t-t'\right) \vec{v}\left(t' \right)dt'=\vec{f}_{\mathrm{r}}\left( t\right) +\vec{f}_{\mathrm{ext}}\left(t\right),   
\label{Combinaison1}
\end{equation}

On the right hand side, $\vec{f}_{\mathrm{ext}}(t)$ represents a
possible external force applied by an operator, for instance with an
optical trap. It should be noted that the expression of the
dissipative force is valid even if the probe is only partially
embedded in the medium, since $\Gamma(t)$ takes into account the
detailed geometry of the contact between the probe and the medium.
   
In the absence of external force, using the Laplace transforms
$\hat{v}(s)=\mathcal{L} \{v(t)\}$, $\hat{f}(s)=\mathcal{L} \{f(t)\}$
and $\hat{\Gamma}(s)=\mathcal{L} \{\Gamma(t)\}$,
equation~\ref{Combinaison1} projected onto one direction becomes:

\begin{equation}
\hat{\Gamma}\left(s\right) \hat{v}\left(s\right)=\hat{f}_{\mathrm{r}}\left(s\right).
\label{Combinaison2}
\end{equation}

Introducing the Laplace transform of the autocorrelation functions for
the probe velocity $\hat{C}_{vv}(s)=\mathcal{L} \{\langle
v(t+\tau)v(t) \rangle_{t}\}$ and for the stochastic forces
$\hat{C}_{ff}(s)=\mathcal{L} \{\langle
f_{\mathrm{r}}(t+\tau)f_{\mathrm{r}}(t) \rangle_{t}\}$, respectively
named velocity spectrum and force power spectrum in the following, we
obtain from equation~\ref{Combinaison2} (see \cite{gallet2009}):

\begin{equation}
\hat{C}_{ff}(s) = \hat{\Gamma}^{2}\left(s\right) \hat{C}_{vv}(s).
\label{Combinaison3} 
\end{equation}    

A measurement of $\hat{C}_{ff}(s)$ through
equation~\eqref{Combinaison3} is not directly accessible in our
experiments. We thus re-express below the right-hand side of
equation~\eqref{Combinaison3} in a more convenient fashion. First, to
measure the viscoelastic response kernel $\Gamma(t)$ defined in
equation~\ref{Combinaison1}, it is required that $f_{\mathrm{r}} \ll
f_{\mathrm{ext}}$. Practically, we will see that $f_{\mathrm{r}}$ is a
few piconewtons, while the force $f_{\mathrm{ext}}$ exerted by optical
tweezers is in the range $10$ to $200 ~\pico\newton$, which
legitimates the approximation. Under these conditions,
equation~\ref{Combinaison1} projected onto the direction of external
force becomes $\int_{0}^{t} \Gamma\left(t-t'\right) \cdot v\left(t'
\right)dt'= f_{\mathrm{ext}}\left(t\right)$. Since the bead's position
can be measured with a much better accuracy than its velocity,  we
decided to measure the creep response function $J(t)$ instead of
$\Gamma(t)$. $J(t)$ is defined by:
\begin{equation}
x(t) = \int_{0}^{t} J \left(t-t'\right) \left.\frac{df_{\mathrm{ext}}}{dt'}\right|_{t'} dt'.
\label{MicrorheologyActive}
\end{equation}
where $f_{\mathrm{ext}}(t)= kx_{\mathrm{f}}(t)$ is the force exerted by the optical trap. After Laplace transforming, equation~\ref{MicrorheologyActive} becomes: 
\begin{equation*}
\hat{J}(s) = \dfrac{\hat{x}(s)}{s \hat{f}_{ext}(s)} =\dfrac{1}{s k}\dfrac{\hat{x}(s)}{\hat{x}_{\mathrm{f}}(s)} 
\label{chapter3:eq:microactive}
\end{equation*}
$\Gamma$ and $J$ are simply related through: $\hat{\Gamma}(s) = \frac{1}{s^{2} \hat{J}(s)}$. 

Furthermore, the velocity spectrum $\hat{C}_{vv}$ is related to the Laplace transform of the mean square displacement $\mathcal{L}\{\Delta x^2 (\tau)\}=\Delta \hat{x}^{2}(s)$ through:  $2\hat{C}_{vv}(s) = s^{2} \Delta \hat{x}^{2}(s)$. 

Combining these different relations, we can rewrite
equation~\ref{Combinaison3} only with observables easily accessible
from passive and active microrheology experiments:
\begin{equation}
\hat{C}_{ff}(s) = \dfrac{\Delta \hat{x}^{2}(s)}{2 s^2 \hat{J}(s)^{2}}.
\label{Combinaison4}
\end{equation}
Thus, to calculate the force power spectrum experienced by a small
probe bound to the cytoskeleton, we need to measure the creep function
$\hat J(s)$ and the MSD $\Delta \hat{x}^{2}(s)$. Ideally, these two
measurements should be simultaneous, but, for practical reasons, they
are successively performed within a short time interval (less than 1
min): the mechanical environment of the probe is assumed to remain
unchanged between the two successive experiments.
\section{Microrheology: effect of ligand density}
\label{Microrhology}

To quantify how the distribution of active forces exerted on the bead is affected by the degree of attachment of the probe to the actin cortex, we modified the RGD concentration during the bead's coating process (see~\ref{Bead coating}). We considered three decreasing RGD concentrations: \textbf{RGD 1/10}, \textbf{\textcolor{blue}{RGD 1/20}} and \textbf{\textcolor{red}{RGD 1/40}}. In this section, all the measurements were performed at room temperature ($\sim25\celsius$)   

\subsection{Passive microrheology results}
\begin{figure*}[h]
\begin{center}
\includegraphics[width=16cm]{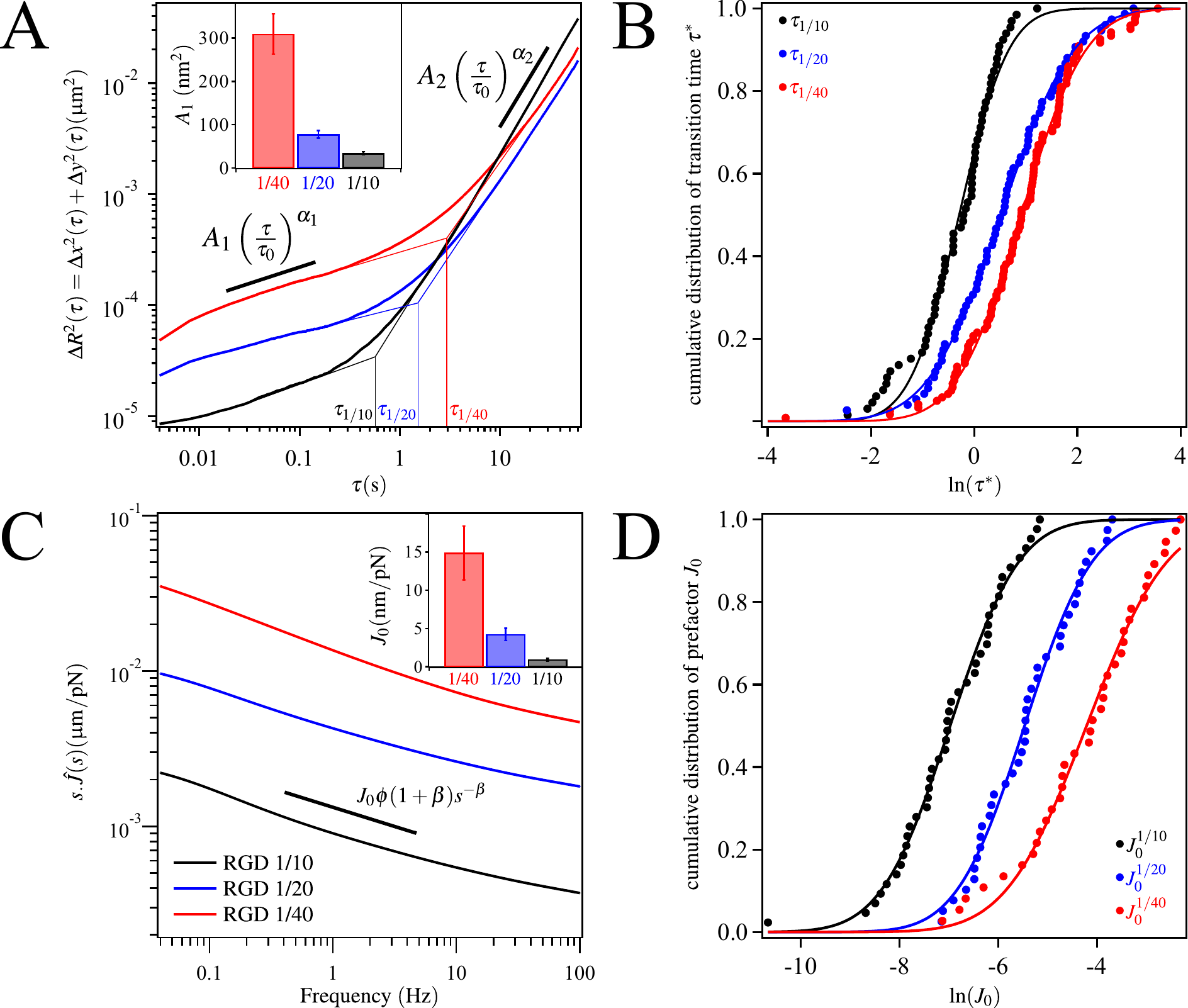}
\end{center}
\caption{ (A) MSD $\Delta R^2(\tau) = \Delta x^2(\tau) + \Delta y^2(\tau)$, geometrically averaged over N realizations, as a function of the time lag $\tau$ for three concentrations of peptide:~RGD 1/10 (black, $N_{1/10}=66$), RGD 1/20 (blue, $N_{1/20}=77$) and RGD 1/40 (red, $N_{1/40}=122$). The times $\tau^*_{1/10}$, $\tau^*_{1/20}$ and $\tau^*_{1/40}$ are the transition times between the subdiffusive and superdiffusive regimes. (B) Cumulative distribution of transition time $\log(\tau^*)$, fitted by a normal cumulative probability function $p(x) = \frac{1}{2} \left(1+ \mathrm{erf} \left(\frac{x-\left\langle x\right\rangle}{\sigma\sqrt{2}} \right)\right)$ with $x = \log(\tau^*)$. (C) Laplace transforms of response functions $s\hat{J}(s)$, geometrically averaged over N realizations, as a function of frequency $s$, for three concentrations of peptide: RGD 1/10 (black, $N_{1/10}=43$), RGD 1/20 (blue, $N_{1/20}=39$) and RGD 1/40 (red, $N_{1/40}=37$). (D) Cumulative distribution of the logarithm of prefoactors $\log(J_{0})$ fitted with with the normal cumulative probability function\label{fig:RGD1}}
\end{figure*}

A probe embedded in a purely viscous liquid in equilibrium adopts a
diffusive motion, its MSD is a linear function of time:
$\langle\Delta{x^{2}_{\mathrm{eq}}}(\tau)\rangle \sim \tau^1$. By
contrast, in a purely elastic medium in equilibrium, the MSD remains
constant $\langle\Delta{x^{2}_{\mathrm{eq}}}(\tau)\rangle \sim
\tau^0$. Now, in a viscoelastic medium in equilibrium, the MSD often
exhibits an intermediate power law behavior:
$\langle\Delta{x^{2}_{\mathrm{eq}}}(\tau)\rangle \sim \tau^\alpha$
with $0<\alpha<1$, and the motion is called subdiffusive. On the other
hand, if the medium is not at thermal equilibrium, partially directed
motion may occur, corresponding to superdiffusive behavior with
$\alpha > 1$. The value $\alpha = 2$ corresponds to ballistic motion.

The results of passive microrheology measurements on beads bound to
the actin cortex are shown in fig~\ref{fig:RGD1}A. The averaged MSD
$\Delta R^2(\tau)=\Delta x^2(\tau)+\Delta y^2(\tau)$ is plotted for
three ligand concentrations.  For each concentration, we observe two
regimes. At short time scale ($\tau \lesssim1~\second$), the motion of
the bead is subdiffusive and can be fitted by a power law: $\Delta
R^2(\tau) \sim A_1 (\tau / \tau_0)^{\alpha_1}$ with $\alpha_1 < 1$;
on the other hand, at long time scale ($\tau \gtrsim 1~\second$), the
motion is superdiffusive: $\Delta R^2(\tau) \sim A_2 (\tau /
\tau_0)^{\alpha_2}$ with $\alpha_2 > 1$. Subdiffusive motion has
already been observed in similar systems
~\cite{Raupach2007,bursac_cytoskeleton_2007,bursac_cytoskeletal_2005,metzner_simple_2007,girard_dictyostelium_2006},
sometimes refered as "caged motion"~\cite{lenormand_directional_2007}:
the probe motion is partially confined by the actin network at short
time scale. By contrast, for $\tau \gtrsim 1~\second$, the probe's
motion is superdiffusive. As we show below, this stems from
non-thermal forces mediated by the cytoskeleton. The origin of such
forces are likely to be molecular motors such as myosins.

One notices that the three curves corresponding to different
concentrations do not overlap. The averaged values of the exponents
and amplitudes of MSD in the subdiffusive and superdiffusive regimes,
calculated from their distribution over $N$ realizations, are reported
in table~\ref{tableau_passive_microrheology} as a function of the
ligand density. In the subdiffusive regime, at short time scale, the
averaged exponent $\overline{\alpha_1}$ does not significantly depend
on the RGD concentration, while the mean amplitude $\overline{A_1}$,
geometrically averaged over all $A_1$ values, increases by one order
of magnitude as the RGD concentration decreases. On the other hand, in
the superdiffusive regime, at longer time scale, the averaged exponent
$\overline{\alpha_2}$ appears to increases slightly when the probe is
more tightly attached to the cortex, while the amplitude
$\overline{A_2}$ decreases. These variations are less pronounced than
in the subdiffusive regime and harder to assess, due to the difficulty
of sampling such late-time regimes. To calculate the numerical value
of $A_1$ and $A_2$ we have set $\tau_0= 1~s$, at the center of the
explored time range.

\begin{table}[h]
\small
  \caption{Averaged values and standard deviations of the exponents and amplitudes of MSD in the subdiffusive and superdiffusive regimes, as a function of the ligand density, calculated from their distributions over N realizations. The crossover time $\overline{\tau^*}$ between the two regimes is also reported. The errors correspond to the standard error (SE) of the mean of our samples. They do not account for the accuracy of the fit to a power-law behaviour.}
  \label{tableau_passive_microrheology}
  \begin{tabular*}{0.48\textwidth}{@{\extracolsep{\fill}}llll}
    		 \hline    		 
    		\small \textbf{RGD conc.} & \small \textbf{1/10} & \small \textbf{1/20} & \small \textbf{1/40} \\    		
    		\hline
    		\small $\overline{\alpha_1}$ & \small 0.32 $\pm$ 0.02 & \small 0.24 $\pm$ 0.01 & \small 0.31 $\pm$ 0.01\\        
    		\small $\overline{A_1}$ ($nm^2$)& \small 34 $\pm$ 4 & \small 78 $\pm$ 8 & \small 301 $\pm$ 46\\    		 
    		\small $\overline{\alpha_2}$  & \small 1.60 $\pm$ 0.03 & \small 1.38 $\pm$ 0.05 & \small 1.19 $\pm$ 0.04\\     		
     		\small $\overline{A_2}$ ($nm^2$) & \small 47 $\pm$ 9 & \small 52 $\pm$ 10 & \small 133 $\pm$ 25 \\ 
	   		\small $\overline{\tau^*}$ ($s$) & \small 0.8 $\pm$ 0.1 & \small 1.7 $\pm$ 0.2 & \small 2.5 $\pm$ 0.2 \\ 
     		\hline  
  \end{tabular*}
\end{table}  

We can define a transition time $\tau^*$ between the two regimes as
the intersection of the two power law fits: $\tau^* = \left(
\frac{A_2}{A_1}\right)^{\frac{1}{\alpha_1 -
    \alpha_2}}$. Figure~\ref{fig:RGD1}B shows the cumulative
distribution of $\log(\tau^*)$, together with the best fit by a
cumulative normal distribution: $p(x) = \frac{1}{2} \left(1+
\mathrm{erf} \left(\frac{x-\left\langle
  x\right\rangle}{\sigma\sqrt{2}} \right)\right)$ (with $x =
\log(\tau^*)$). For each RGD concentration, the transition time
$\tau^*$ is log-normally distributed. Note that $\overline{\tau^*}$
shifts towards longer times as the concentration of RGD decreases,
which likely reflects the fact that the probe is more weakly attached
to the cell.

\subsection{Active microrheology results}
\label{Active microrheology results}
The results of active microrheology measurements are shown in fig~\ref{fig:RGD1}C. The response function $s\hat{J}(s)$, geometrically averaged over N realizations, is drawn as a function of frequency $s$ for three ligand densities. As shown in previous works ~\cite{lenormand_linearity_2004,desprat_creep_2005,balland_power_2006,Hoffman2006,icard-arcizet_cell_2008}, $s\hat{J}(s)$ behaves as a power law of $s$, and may be fitted by: $s\hat{J}(s) = \hat{J_0} s^{-\beta}$. Taking the inverse Laplace transform, this leads to a power law dependance of the time response function $J(t) = J_0 t^{\beta}$, with $J_0 = \frac{\hat{J_0}}{\phi(1+\beta)}$ ($\phi$ is the Euler function). The results, averaged over N realizations, are reported in table~\ref{tableau_active_microrheology} as a function of the ligand density. No significant variation of the exponent $\overline{\beta}$ is observed with RGD concentration: $\overline{\beta}$ remains close to $\approx 0.2$, a value similar to those already reported in the literature, from different experimental techniques and for different cell types. 

\begin{table}[h]
\small
  \caption{Averaged values and standard deviations of the exponents and amplitudes of the active response function $J(t)$, as a function of the ligand density, calculated from their distributions over N realizations.}
  \label{tableau_active_microrheology}
  \begin{tabular*}{0.48\textwidth}{@{\extracolsep{\fill}}llll}
 			\hline    		 
    		\small \textbf{RGD conc.} & \small \textbf{1/10} & \small \textbf{1/20} & \small \textbf{1/40} \\    		
    		\hline
    		\small $\overline{\beta}$ & \small 0.20 $\pm$ 0.02 & \small 0.19 $\pm$ 0.02 & \small 0.25 $\pm$ 0.03\\       
    		\small $\overline{J_0}$ ($nm/pN$)& \small 0.9 $\pm$ 0.2 & \small 4.2 $\pm$ 0.8 & \small 14.9 $\pm$ 3.5\\    		 	\hline  
  \end{tabular*}
\end{table} 
 
On the other hand, we observe that the amplitude $\hat{J}_0$ of $\hat{J}(s)$ is RGD-dependent, $\hat{J}(s)$ (and thus $J(t)$) increases when the binding between the bead and the cortex is weaker. The distribution of the prefactor $J_0$ of $J(t)$ is shown in figure~\ref{fig:RGD1}D. It is log-normally distributed, as already pointed out in~\cite{desprat_creep_2005,icard-arcizet_cell_2008}. Its average value $\overline{J_0}$ significantly depends on the ligand density (see table~\ref{tableau_active_microrheology}). As expected, we observe that the tighter the beads are attached to the cortex, the harder it is to move them. Consistently with~\cite{bursac_cytoskeleton_2007,icard_arcizet_modifications_2007}, the $\{$bead+cortex$\}$ compliance decreases while the ligand density increases.

\subsection{Modelling the origin of the bead's active motion}
\label{Modelling the origin of the bead's active motion}

\subsubsection{Building up the model}
\begin{figure}[h]
  \begin{center}
\includegraphics[width=\columnwidth]{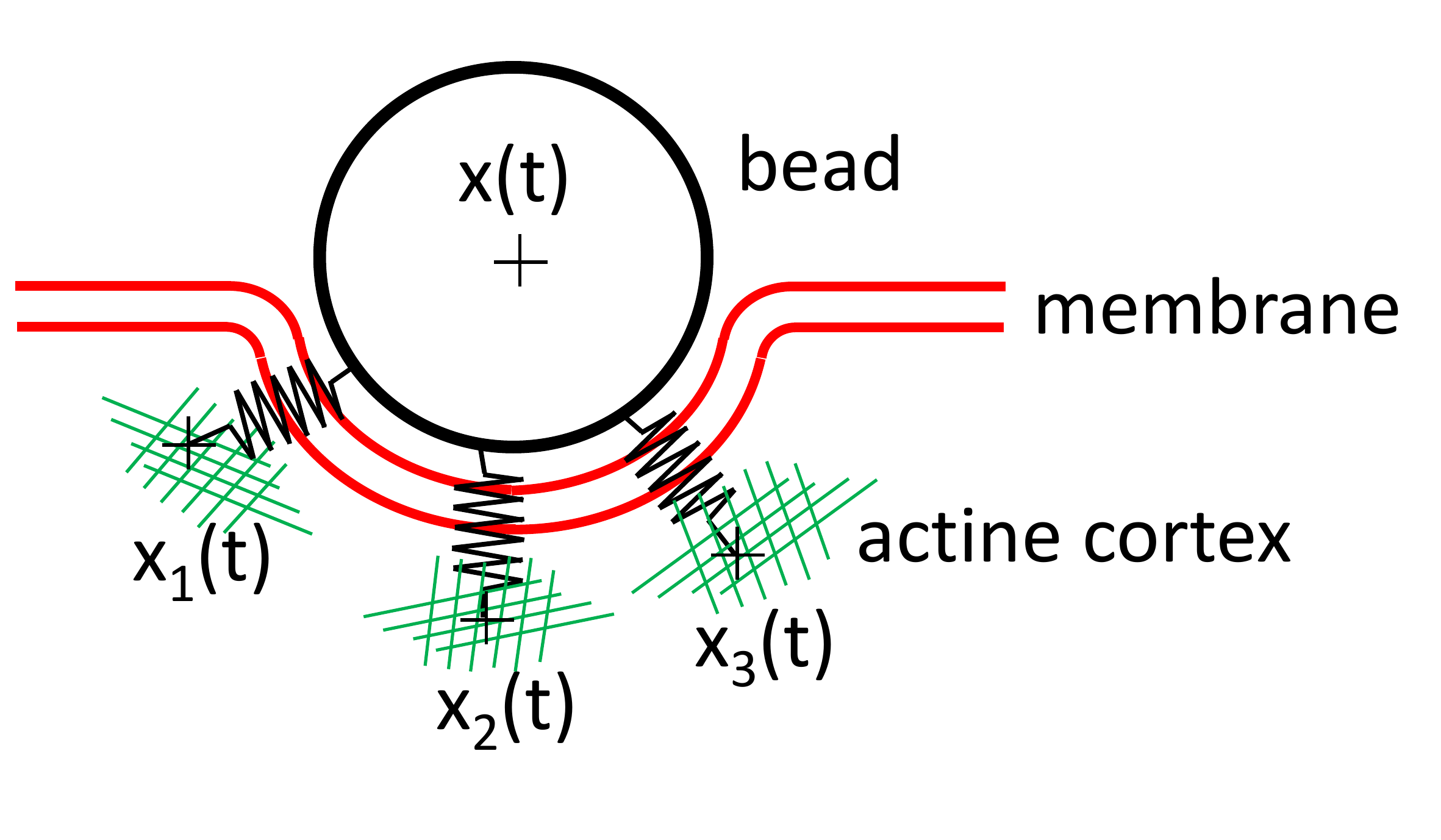}
  \end{center}
  \caption{Schematic picture of the bead connected to the cytoskeleton through $N$ elastic RGD bonds of stiffness $k$. $x(t)$ describes the position of the bead while the $x_i(t)$ describe the positions of the anchors, which undergo activated viscoelastic dynamics, due to the forces exerted by the motors on the cytoskeleton.}\label{fig:scheme} 
\end{figure}

In this section we develop a simple model to account for the different regimes observed during the spontaneous motion of the bead at different time scales and different ligand concentrations. In particular, we want to rationalize the dependence of the MSD and of the creep function on time and ligand concentration. To do so, we model the bead by its position $x(t)$. The connection between the bead and the cytoskeleton is then modeled by $N$
RGD bonds anchored to the cytoskeleton and behaving as harmonic springs of stiffness $k$ (see Fig~\ref{fig:scheme}).
We refer to the corresponding locations $x_i(t)$ as the ``active anchors''. As a minimal model we write for the dynamics of the bead:
\begin{equation}\label{dyn:x}
  \gamma \dot x = -\sum_i k_i(x-x_i) + \sqrt{2 \gamma T} \eta\;,
\end{equation}
where  $\eta$ is a Gaussian white noise of correlations
\begin{equation}
  \langle \eta(t) \eta(t') \rangle = \delta(t-t')\;,
\end{equation}
that represents thermal fluctuations. The fluid surrounding the bead is here assumed to be purely viscous. On the contrary, the $N$ active
anchors evolve according to visco-elastic generalized Langevin equations
\begin{equation}\label{dyn:probe}
  \int_{-\infty}^{t}\gamma_i\left(t-t'\right) \dot x_i\left(t' \right)dt'= \sum_i k(x(t)-x_i(t)) +  \xi_i(t) + f_i^{\rm a}(t)\;.
\end{equation}
The $\gamma_i$ are damping kernels that model the viscoelasticity of
the cytoskeleton, the noises $\xi_i$ are thermal fluctuations that
affect the active anchors. They thus satisfy a generalized
Stokes-Einstein relation with respect to the damping kernels
$\gamma_i$, which reads, in Laplace space,
\begin{equation}
  \langle \hat\xi_i(s)\hat\xi_j(s')\rangle =  \delta_{ij} T  \frac{\hat\gamma_i(s)+\hat\gamma_i(s')}{s+s'}\;.
\end{equation}
The forces $f_i^{\rm a}$ are active forces that are generated by
molecular motors and transmitted to the active anchors through the
cytoskeleton. We will later on assume that they are strongly
correlated spatially. These forces are also correlated in time, but
need not (and will not in general) satisfy a FDT relation with respect
to the $\gamma_i$.

\subsubsection{Effective dynamics of the bead}

In the Laplace variable $s$ conjugate to time, the dynamics of the
active anchors can be solved as
\begin{equation}
\hat x_i(s) = \frac{k \hat x(s) + \hat \xi_i(s)-\hat \gamma_i(s) x_i^0}{k+s \hat \gamma_i(s)}\;,
\end{equation}
where we have introduced initial positions $x_i^0$ for the
anchors. The bead's position is then given in Laplace space by
\begin{equation}\label{eq:dynx}
  \hat x(s) s \hat \Gamma(s) = \sqrt{2 \gamma T} \hat \eta(s) + \sum_{i=1}^N [\hat \xi_i(s)+\hat f_i^a(s)-\hat \gamma_i(s) x_i^0]\;,
\end{equation}
where we have introduced the effective damping felt by the bead
\begin{equation}
  \hat \Gamma(s) \equiv \gamma + \sum_i \frac{\hat \gamma_i}{1+\frac{s}{\hat\omega_i}}\;.
\end{equation}
Here, $\hat \omega_i(s)\equiv k/\hat\gamma_i$ is the inverse
relaxation times of the RGD molecules.

Note that the interactions between the bead and the cytoskeleton dress up the viscous damping $\gamma$ into an effectively visco-elastico one $\Gamma$. In practice, for the experiments, we
assume that the RGD bonds are very stiff, and that their relaxation time are much shorter than any measured time-scales, so
\begin{equation}\label{eq:Gamma}
  \frac{s}{\hat \omega_i} \ll 1 \qquad\mbox{and hence}\qquad  \hat \Gamma(s) \simeq \gamma + \sum_i \hat \gamma_i(s) \simeq N \hat \gamma_{\rm c}(s)
\end{equation}
where we have finally assumed that most of the damping comes from the
anchoring to the cytoskeleton and we have introduced $\hat \gamma_{\rm c}$
a representative damping kernel of an active anchor. 

\subsubsection{The effective damping}
At this stage, we can already comment on the experimental measurements of $\hat J(s)$. In Fig.~\ref{fig:RGD1}C, we see that $s\hat J(s)$ is well
described by a power-law $s\hat J(s)=\hat{J_0}s^{-\beta}$, with the amplitude $\hat{J_0}$ that decreases as the concentration of RGD increases and $\beta$ which is roughly independent of $N$. Using that
\begin{equation}
  \hat J(s) = \frac1{s^2 \hat \Gamma(s)}
\end{equation}
one gets that
\begin{equation}
  \hat \Gamma(s) \sim \frac 1 {\hat{J_0} s^{1-\beta}}
\end{equation}
This is qualitatively consistent with equation~\eqref{eq:Gamma} that predicts $\hat{J_0}\sim 1/N$. From the experiments, we also get $\beta\simeq 0.2$ (see Table~\ref{tableau_active_microrheology}) so
that
\begin{equation}\label{eq:scalinggamma}
  \Gamma(t) \sim N  {t^{-\beta}} \sim N {t^{-0.2}}
\end{equation}
Note that these scalings allow us to characterize the viscoelasticity experienced by the active anchors through: $\hat \gamma_c(s)\sim
s^{\beta-1}$ and $\gamma_c(t) \sim t^{-\beta}$. Using our model we have thus already inferred rheological properties of the cytoskeleton, which will also prove useful in the analysis of the bead's MSD.

\subsubsection{The Mean-Square Displacement}
\label{sec:MSDscaling}
Eq.~\eqref{eq:dynx} directly leads to an expression for $\langle \hat
x(s) \hat x(s') \rangle$:
\begin{align}
  \langle \hat x(s) \hat x(s') \rangle &=& \frac{2 \gamma T}{s s'(s+s') \hat\Gamma(s) \hat\Gamma(s')} + \frac{\sum_i  T [\hat \gamma_i(s)+ \hat \gamma_i(s')]}{s s' (s+s') \hat\Gamma(s) \hat\Gamma(s')}\nonumber\\&& +\frac{\sum_{i,j} \hat\gamma_i \hat \gamma_j \langle x_i^0 x_j^0 \rangle}{s s' \hat\Gamma(s) \hat\Gamma(s')}+ \frac{N^2[\hat C_{f_af_a}(s)+\hat C_{f_af_a}(s')]}{s s' (s+s') \hat\Gamma(s) \hat\Gamma(s')}\label{eq:MSD}
\end{align}
Here we have assumed that all the active anchors are experiencing the same active force $f_i^{\rm a}=f^{\rm a}$, which is thus correlated over a scale comparable to the bead's size, and we have introduced $\hat C_{f_af_a}(s)$ which is the Laplace transform of $C_{f_af_a}(t)=\langle f^{\rm a}(t)f^{\rm a}(0)\rangle$. We now want to disentangle the contributions of each terms in Eq.~\eqref{eq:MSD} to identify the various regimes of the MSD presented in Fig.~\ref{fig:RGD1}A.

Using Bromwich formula, the MSD can be obtained from
\begin{equation}\label{eq:int1}
  \langle x(t)^2 \rangle = -\frac {1}{4\pi^2} \int ds ds' e^{s t} e^{s' t}   \langle \hat x(s) \hat x(s') \rangle
\end{equation}
where the integrals are carried parallel to the imaginary axis, on the right of any singularities of $\langle \hat x(s) \hat x(s')\rangle$.  Making a change of variable from $s$ to $u=s t$, Eq.~\eqref{eq:int1} becomes 
\begin{equation}\label{eq:int2}
  \langle x(t)^2 \rangle= -  \frac {1}{4\pi^2 t^2} \int du du' e^{u} e^{u'}   \langle \hat x(u/t) \hat x(u'/t) \rangle
\end{equation}
The scaling of $\langle x(t)^2 \rangle$ with $t$ is thus the same as that of $t^{-2} \langle \hat x(1/t) \hat x(1/t) \rangle$. 

Using the scaling~\eqref{eq:scalinggamma} deduced from the
experimental data, we can now assess the contributions to the MSD of
each terms entering Eq.~\eqref{eq:MSD}. The first term in the
r.h.s. of~\eqref{eq:MSD} stems from thermal noise directly acting on
the bead, it leads to a contribution scaling as
\begin{equation}
  \frac{2 \gamma T}{s s'(s+s') \hat\Gamma(s) \hat\Gamma(s')} \to \frac{2 \gamma t ^{2 \beta -1 }}{N^2}\;.
\end{equation}
Interestingly, it gives a vanishing contribution to the MSD at large
times. The sole injection of energy through the beads cannot overcome
the viscoelastic damping of the cytoskeletton.

The second term in~\eqref{eq:MSD} comes from the thermal forces acting
on the cytoskeletton, it leads to the following contribution to the MSD:
\begin{equation}
  \frac{\sum_i  T [\hat \gamma_i(s)+ \hat \gamma_i(s')]}{s s' (s+s') \hat\Gamma(s) \hat\Gamma(s')}  \to \frac 1 N t^\beta\label{eq:thermal}
\end{equation}
This regime is perfectly compatible with the thermal regime reported at short times in Fig.~\ref{fig:RGD1}A, regarding both the values of the exponents (within experimental accuracy) and the dependence of the amplitude $A_1$ on the number of RGD bonds $N$.

The third term comes from the correlations the initial positions of the RGD bonds, it leads to
\begin{equation}
  \frac{\sum_{i,j} \hat\gamma_i \hat \gamma_j \langle x_i^0 x_j^0 \rangle}{s s' \hat\Gamma(s) \hat\Gamma(s')}\sim
  \left\{\begin{array}{l l}
    {\cal O}(1) & \mbox{if $x_i^0$ and $x_j^0$ are correlated}\\
    {\cal O}(\frac 1 N ) & \mbox{if $x_i^0$ and $x_j^0$ are uncorrelated}    
  \end{array}\right.
\end{equation}
Unsurprisingly, in both cases, this initial contribution does not impact the late-time regime.

The final term stems from the active forces
\begin{equation}
  \frac{N^2[\hat C_{f_af_a}(s)+\hat C_{f_af_a}(s')]}{s s' (s+s') \hat\Gamma(s) \hat\Gamma(s')} \to t^{2 \beta -1} \hat C_{f_af_a}\Big(\frac 1 t\Big)\label{eq:active}
\end{equation}
To get an exponent $\alpha_2=1.2-1.6$ compatible with the experimental
values reported in table~\ref{tableau_passive_microrheology}, one thus
needs $\hat C_{f_af_a}(\frac 1 t)\sim t^{1.8-2.2}$ and $C_{f_af_a}(t)\sim t^{0.8-1.2}$. The fact that the autocorrelation function of active forces $C_{f_af_a}(t)$ might increase with time should not be a surprise: such a behavior has been previously described \cite{robert_vivo_2010,gallet2009,mizuno_high-resolution_2009}, and means that the whole actin cortex surrounding the bead undergoes persistent and partially directed forces over time scales larger than $\sim 1~min$. This will be confirmed by the experimental results in section \ref{Force power spectrum}.

The two regimes observed in the experiments thus arise from the second and last terms in~\eqref{eq:MSD}, which correspond to the thermal and active fluctuations of the cytoskeletton, transmitted to the bead through the RGD bonds. The crossover between these two regimes occurs
at a time $\tau^*$ obtained by balancing~\eqref{eq:thermal} and~\eqref{eq:active}
\begin{equation}
  \frac 1 N (\tau^*)^\beta = (\tau^*)^{\alpha_2} \to  (\tau^*) \sim \left(\frac 1 N \right)^{\frac 1 {\alpha_2-\beta}} 
\end{equation}
This explains the trend reported in the last line of table~\ref{tableau_passive_microrheology}: $\tau^*$ decreases with $N$.

\section{Force power spectrum and distance from equilibrium}
\label{Force power spectrum and distance from equilibrium}

\subsection{Definition of the effective temperature}

In thermal equilibrium, the fluctuation-dissipation theorem applies. In this case, the force spectrum $\hat{C}_{ff}^{eq}$ only depends on the bath temperature, according to ~\cite{mason_optical_1995}:

\begin{equation}
\hat{C}_{ff}^{eq}(s)=\dfrac{k_B T}{s^2 \hat J(s)}.
\label{spectre_equilibre}
\end{equation}

In an out-of-equilibrium system, equation~\ref{spectre_equilibre} does not hold. Using a formal analogy, one can define a frequency-dependent effective temperature $T_{\mathrm{eff}}$, as:

\begin{equation}
 \hat{C}_{ff}(s)=\dfrac{\Delta \hat{x}^{2}(s)}{2 s^2 \hat{J}(s)^{2}}=\dfrac{k_B T_{\mathrm{eff}}(s)}{s^2 \hat J(s)}.
\end{equation}

Alternatively, one can define a function $\theta(s)$ measuring the distance from thermal equilibrium for the force power spectrum:
\begin{equation}
 \theta(s) = \dfrac{\hat{C}_{ff}(s)}{\hat{C}_{ff}^{eq}(s)}=\dfrac{T_{\mathrm{eff}}(s)}{T}=\dfrac{\Delta \hat{x}^{2}(s)}{2 k_B T \hat{J}(s)}.
 \label{distance}
\end{equation}

In equilibrium $\theta(s)=1$, and the stochastic forces acting on the probe reduce to thermal forces. When $\theta(s)>1$, active forces $F_{\mathrm{a}}(t)$ generated by biological activity also contribute to the force power spectrum, bringing the system out-of-equilibrium.

\subsection{Force power spectrum}
\label{Force power spectrum}
\begin{figure*}[h]
\begin{center}
\includegraphics[width=\textwidth]{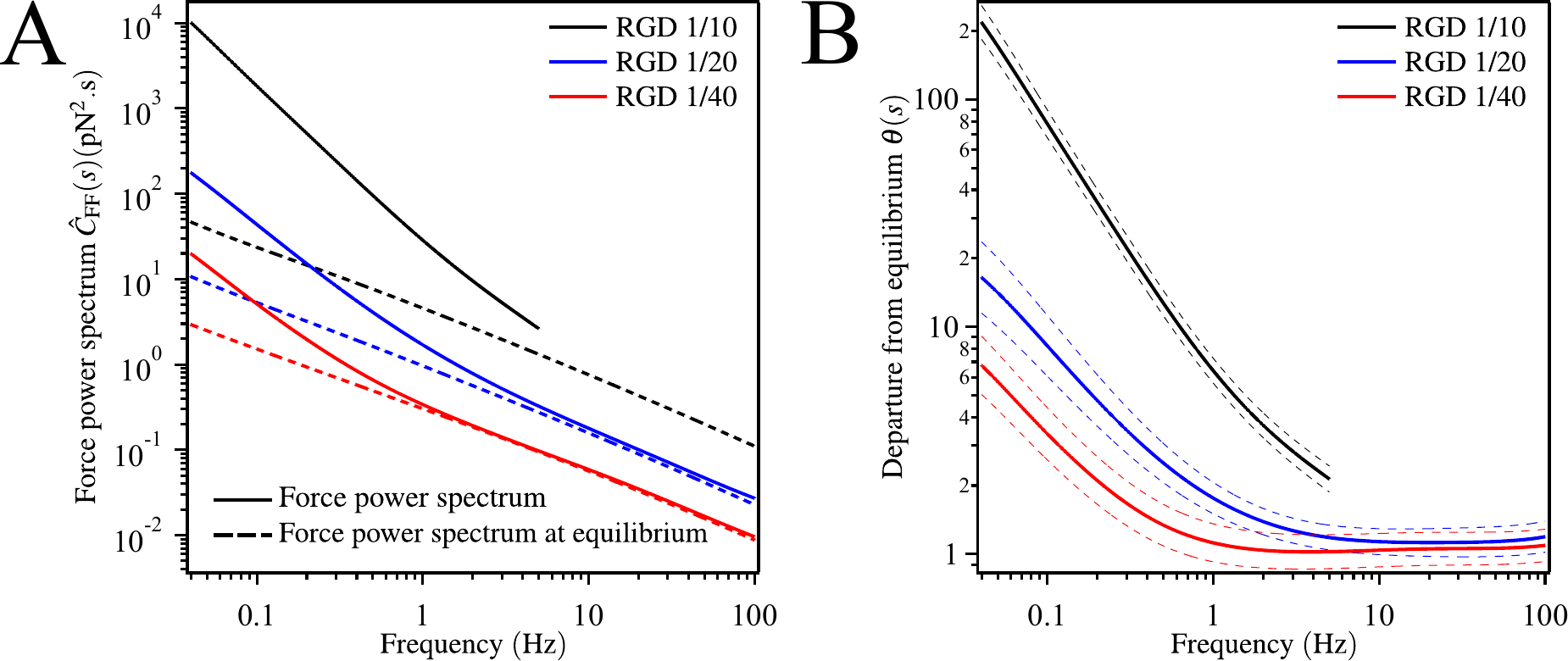}
\end{center}
\caption{(A) Comparison between the geometrical mean of the real force power spectrum $\hat{C}_{ff}(s)$ experienced by the bead, calculated using equation ~\ref{Combinaison4} (solid line), and the force power spectrum $\hat{C}_{ff}^{eq}$, calculated from equation~\ref{spectre_equilibre}, as if the fluctuation-dissipation theorem in equilibrium would apply (dashed lines). (B) The mean distance from equilibrium $\hat{\theta}(s)$ (equation~\ref{distance}) is drawn as a function of frequency $s$ for different RGD concentrations (red: RGD 1/40, blue: RGD 1/20 and black: RGD 1/10). Dashed lines represent the extent from mean value, $\pm$ SE, with $N_{1/10}=43$, $N_{1/20}=39$ and $N_{1/40}=37$.\label{fig:RGD2}}
\end{figure*}

According to equation~\ref{Combinaison4}, we can retrieve the power spectrum $\hat{C}_{ff}(s)$ of the stochastic forces exerted on the probe, by combining active and passive microrheology measurements performed on the same probe. Independently, by using equation~\ref{spectre_equilibre}, we can predict what would the power spectrum $\hat{C}_{ff}^{eq}$  be if the system were in equilibrium. Taking the ratio of these two quantities, we have access to the distance from equilibrium $\hat{\theta}(s)$ (eq~\ref{distance}). 

The power spectrum $\hat{C}_{ff}$ is plotted in figure~\ref{fig:RGD2}A, together with the power spectrum $\hat{C}_{ff}^{eq}$ in equilibrium (solid lines and dashed lines, respectively), for the three ligand densities. $\hat{C}_{ff}^{eq}$ behaves as a power law over the full frequency range, in accordance with the behavior of the viscoelastic response $\hat{J}(s)$ (eq~\ref{spectre_equilibre}). The frequency dependence of the force power spectrum in equilibrium is related to the delayed memory kernel in the linear response of the material. 

The solid lines in figure~\ref{fig:RGD2}A represent the mean power spectrum $\hat{C}_{ff}(s)$ of the forces actually experienced by the bead. One can identify two regimes, corresponding to the subdiffusive and superdiffusive behavior of the bead trajectory. At high frequency ($s > 1$Hz) or short time scale ($t < 1\second$), $\hat{C}_{ff}(s)$ has the same amplitude and frequency dependence as the spectrum calculated in equilibrium, and the distance from equilibrium $\hat{\theta}(s)$ is equal to 1 (figure~\ref{fig:RGD2}B). In this range, the behavior is identical to thermal equilibrium one, and we conclude that the thermal forces $\xi(t)$ bring a dominant contribution to the total amplitude of the stochastic forces $f_r(t)$, consistently with the model's predictions.  

Oppositely, in the the low frequency regime ($s < 1$ Hz or $t > 1~\second$) the measured force spectrum $\hat{C}_{ff}(s)$ and the calculated spectrum in equilibrium $\hat{C}_{ff}^{eq}(s)$ are quite different. The distance from equilibrium $\hat{\theta}(s)$ is frequency-dependant and increases at small $s$ (large time), up to $\approx 100$. Thus, in this regime, active forces $f_a(t)$ generated in the actin cortex are dominant over thermal forces $\xi(t)$. Furthermore, $\hat{\theta}(s)$ is found to increase with the coating concentration. This indicates that the bead experiences higher active forces when it is more tightly bound to the actin cortex.

In this time range ($t > 1~\second$ or $s < 1~\hertz $), we fitted
$\hat{C}_{ff}$ by a power law: $\hat{C}_{ff}(s) = c' (s/s_1)
^{-\gamma-1}$, from which we retrieve $\langle f_r(t+\tau) f_r(t)
\rangle_t = c (\tau / \tau_1)^\gamma $ with $ \tau_1= 1/s_1$ and $c =
\frac{c'}{\tau_1 \phi(\gamma+1)}$, $\phi$ being the Euler
function. The prefactor $c$ represents the amplitude of the
autocorrelation function of the active forces exerted on the
probe. Chosing $\tau_1 = 10\second$ in the middle of the range of
interest, we find $c_{1/10} > c_{1/20} > c_{1/40}$, in qualitative
agreement with our model (see table~\ref{tableau_force_spectrum}).

\begin{table}[h]
\small
  \caption{Amplitude and exponent of the autocorrelation function of the stochastic forces exerted on the probe: $\langle f_r(t+\tau) f_r(t) \rangle_t = c (\tau / \tau_1)^\gamma $, as a function of the ligand density.}
  \label{tableau_force_spectrum}
  \begin{tabular*}{0.48\textwidth}{@{\extracolsep{\fill}}llll}
 			\hline    		 
    		\small \textbf{RGD conc.} & \small \textbf{1/10} & \small \textbf{1/20} & \small \textbf{1/40} \\    		
    		\hline
    		\small $c$ $(pN^2)$ & \small 206 & \small 4.8 & \small 0.6 \\       		
    		\small $\gamma$ & \small $0.9$  & \small $\gtrsim 0.5$ & \small $\gtrsim 0.5$\\    		 
    		\hline 
  \end{tabular*}
\end{table} 
 
Note that the exponent $\gamma$ also seems to depend on the ligand density: $\gamma_{1/10} > \gamma_{1/20} \approx \gamma_{1/40}$, although the narrow frequency range accessible for RGD 1/20 and RGD 1/40 does not allow an accurate measurement of $\gamma$. Therefore it is hard to decide to what extent this difference is significant or not. 

\subsection{Discussion}
\label{Discussion}

The value of $c^{1/2}$ has the dimension of a force, and leads to an order of magnitude of the average force applied to the bead at the scale $\tau_1 = 10\second$. It is enlightening to compare this amplitude to the force exerted by a single myosin \textit{in vitro}, about $3-4~\pico\newton$~\cite{finer_single_1994}. This result is consistent with the assumption that the active forces are generated by a few molecular motors pulling on the actin filament network surrounding the probe.

Moreover, the variations of $c$ with the ligand concentration indicate that the average force amplitude increases with the number of links between the network and the probe, and consequently with the average number of active motors pulling simultaneously on the bead.

The fact that $\gamma >0$, and that the autocorrelation function $C_{ff}(t)$ increases with time (as predicted in section \ref{sec:MSDscaling}) should not be surprising. Indeed, this resembles to the retrograd flow of the actin network at the back of the cell, which has been profusely reported in the literature (see for instance \cite{Choquet_1997,Allioux_2009,Pollard_Borisy_2003,Alexandrova_Meister_2008}), as due to the cooperative action of molecular motors pulling on the cortex: the persistence time scale of this directed motion is over our experimental time window $\sim 100~s$, which explains why we do not observe any decay of the autocorrelation function $C_{ff}(t)$ on our observation time.

A possible interpretation of the fact that $\gamma$ increases with the ligand density could be that a stronger binding between the bead and the network leads to a higher structuration of the filaments around the bead, and favors the coordination between the motors: this is expected to reinforce the directionnality of the motion, consistently with the model proposed by Robert \textit{et al.}~\cite{robert_vivo_2010,magjarevic_numerical_2010}.

\begin{figure}[h!]
\begin{center}
\includegraphics[width=\columnwidth]{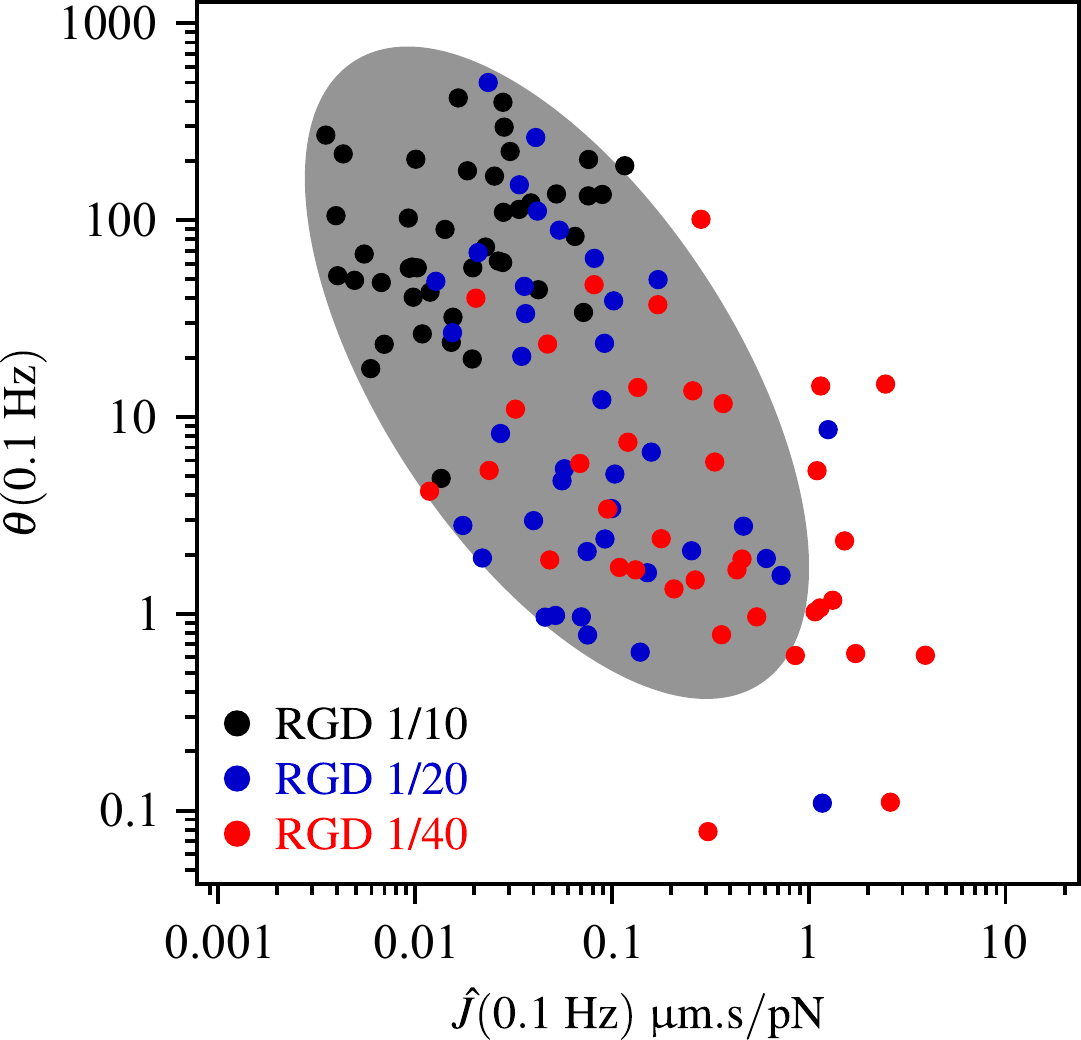}
\end{center}
\caption{Scatter plot of the distance from equilibrium $\theta(s_1)$ versus response function value $\hat{J}(s_1)$, at $s_1 = 0.1~\hertz$ ($\tau_1 = 10~\second$), measured at room temperature and for three concentrations of peptide: RGD 1/10 (black), RGD 1/20 (blue) and RGD 1/40 (red). The ellipse is drawn according to a principal component analysis on the logarithm of $\hat{J}(s_1)$ and $\theta(s_1)$, after removing the $10\%$ most distant points from the average. \label{fig:J_vs_Teta_RGD}}
\end{figure}

Another argument conforting this interpretation comes from the obvious correlation between the response function $\hat{J}(s)$ and the distance from equilibrium $\theta(s)$ in the out-of-equilibrium frequency range ($s < 1Hz$). Indeed, in figure~\ref{fig:J_vs_Teta_RGD}, $\theta(s)$ is plotted versus $\hat{J}(s)$ for all cells and all ligand densities, at a given frequency $s_1=0.1~\hertz$ chosen in the range of interest. Both $\theta(s_1)$ and $\hat{J}(s_1)$ vary over more than three orders of magnitude, but one notices that, on average, $\theta(s_1)$ is a decreasing function of $\hat{J}(s_1)$. As a guide for the eye, the grey ellipse is drawn according to a principal component analysis on the logarithm of $\hat{J}(s_1)$ and $\theta(s_1)$, after removing the $10\%$ most distant points from the average point. Independantly of the ligand density, one can draw through the cloud of points a power law relationship between these two observables. This means that the bead-cortex system is further from equilibrium when the link between the bead and the actin network is tighter: a higher rigidity is correlated to higher pulling forces.

\section{Modifications of metabolism}
\subsection{Influence of temperature}
\begin{figure*}[h!]
\begin{center}
\includegraphics[width=\textwidth]{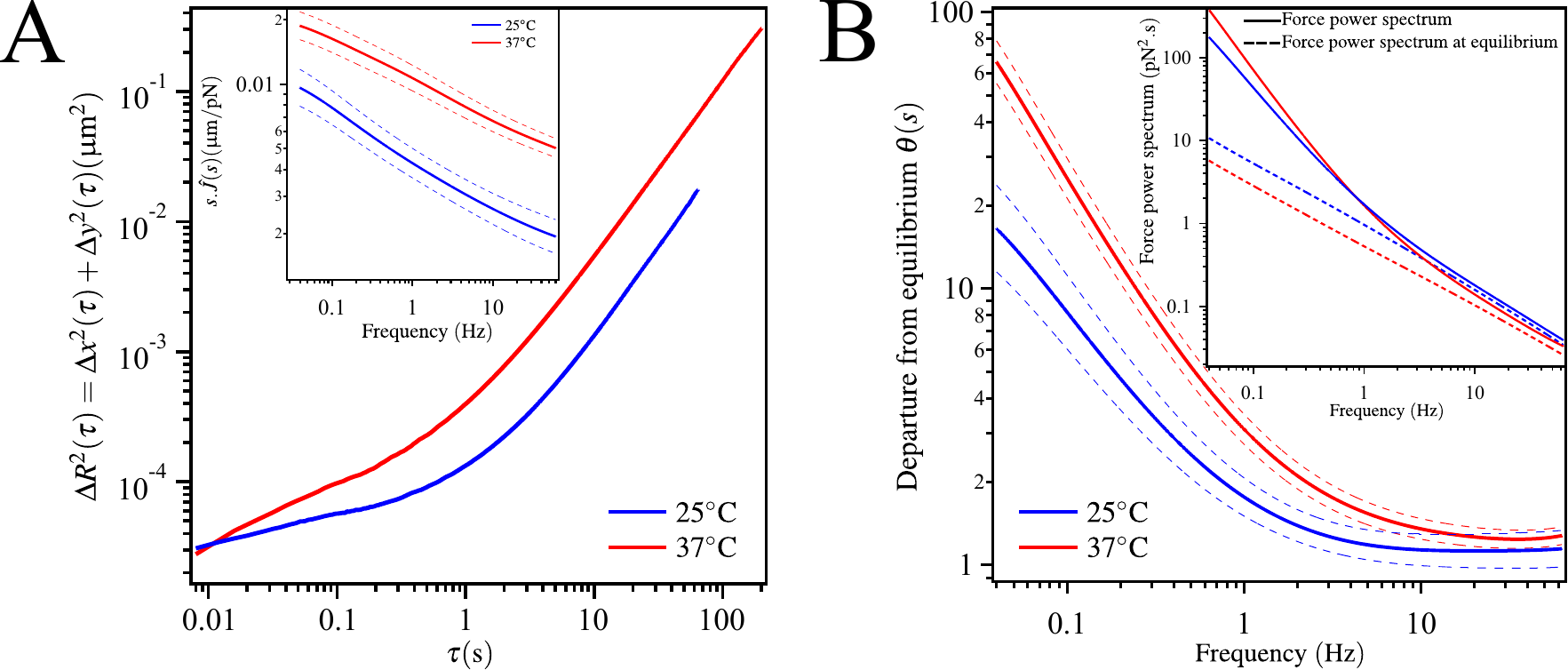}
\end{center}
\caption{Comparison of active and passive measurements at $25\celsius$ (blue), and $37\celsius$ (red). (A) MSD $\Delta R^2(\tau) = \Delta x^2(\tau) + \Delta y^2(\tau)$; inset of (A): Laplace transforms of response functions $\hat{J}(s)$. (B) Mean distance from equilibrium $\hat{\theta}(s)$; inset of (B): force power spectrum $\hat{C}_{ff}(s)$. The data are geometrically averaged over $N=39$ realizations at $25\celsius$  and $N=124$ realizations at $37\celsius$ \label{fig:Celsius}}
\end{figure*}

\subsubsection{Results}
To probe the influence of temperature on the amplitude of active forces, we compared the results of active and passive microrheology measurements on beads coated with the same peptide concentration RGD 1/20, but at two different temperature: $25\celsius$ and $37\celsius$.

The response function $s\hat{J}(s)$, geometrically averaged over N realizations, is plotted in the inset of figure~\ref{fig:Celsius}A (blue curve: $T=25\celsius$ and $N=39$; red curve: $T=37\celsius$ and $N=124$). As mentioned in section \ref{Active microrheology results}, $s\hat{J}(s)$  behaves as a power law of $s$, and thus $J(t) = J_0 t^{\beta}$. Both the averaged amplitude $\overline{J_0}$ and exponent $\overline{\beta}$ depend on the temperature (see table~\ref{tableau_temperature}). This indicates that the bead's environment is more compliant and more dissipative at physiological temperature $T=37\celsius$ than at room temperature. 

\begin{table}[h]
\small
  \caption{Evolution of the parameters describing the response function, the MSD and the force correlation function between $25\celsius$ and $37\celsius$.}
  \label{tableau_temperature}
  \begin{tabular*}{0.48\textwidth}{@{\extracolsep{\fill}}lll}
 			\hline    		 
    		\small \textbf{Temperature} & \textbf{$25\celsius$} & \textbf{$37\celsius$} \\    			
    		\hline
 			$\overline{\beta}$ & 0.19 $\pm$ 0.02 &  0.29 $\pm$ 0.02\\       		
			$\overline{J_0}$ ($nm/pN$)&  4.2 $\pm$ 0.8 &  7.9 $\pm$ 1.2\\    		 
    		$\overline{A_1}$ ($nm^2$) & $78\pm 10$  & $238\pm 22 $\\
    		$\overline{\alpha_1}$  & $0.24\pm 0.01$  & $0.44 \pm 0.01$ \\
    		$\overline{A_2}$ ($nm^2$) & $52\pm 10$  & $188\pm 19$  \\
    		$\overline{\alpha_2}$  & $1.38\pm 0.05$  & $1.44\pm 0.02$ \\
    		$\overline{\tau^*}$ ($s$) & $1.7\pm 0.2$  & $1.3\pm 0.1$ \\
			$c$ $(pN^2)$ &  4.8 &  7.6\\       		
    		$\gamma$ & $\approx 0.5$  &  $0.8$\\    		 
    		\hline  
  \end{tabular*}
\end{table} 
  
Results of passive measurements at $25\celsius$ and $37\celsius$ are shown on figure~\ref{fig:Celsius}A. At both temperatures, the MSD follows a subdiffusive regime for $\tau \lesssim 1 \second$ and a superdiffusive one for $\tau \gtrsim 1 \second$. In the full time range, the amplitude $\Delta x^2(\tau)$ at $37\celsius$ is larger than at $25\celsius$. The results for the prefactors $A_1$ and $A_2$ and the exponents $\alpha_1$ and $\alpha_2$ for $\Delta R^2(\tau)$ in the subdiffusive and superdiffusive regime (see \ref{Passive microrheology}) are shown in table~\ref{tableau_temperature}, respectively averaged over $N_{25\celsius} = 77$ and $N_{37\celsius} = 156$ realizations.

For both regimes, the prefactors and exponents are higher at $37\celsius$ than at $25\celsius$. In the superdiffusive regime, the average displacement appears more directed at physiological temperature. This dependance of $\alpha_2$ has already been pointed out by Bursac \textit{et al.}~\cite{bursac_cytoskeletal_2005}. We also measured the transition time $\tau^*$ between subdiffusive and superdiffusive regime: $\overline{\tau^*}_{37\celsius} < \ \overline{\tau^*}_{25\celsius}$.

As in section~\ref{Force power spectrum}, we combined passive and active microrheology measurements to get the total force power spectrum $\hat{C}_{ff}(s)$, the force power spectrum in equilibrium $\hat{C}_{ff}^{eq}(s)$ (respectively solid and dashed lines in the inset of figure~\ref{fig:Celsius}B), and the distance from thermal equilibrium $\theta(s)  = \hat{C}_{ff}(s) / \hat{C}_{ff}^{eq}(s)$ (figure~\ref{fig:Celsius}B), at $25\celsius$ (red lines) and $37\celsius$ (blue lines). The force power spectra $\hat{C}_{ff}(s)$ and $\hat{C}_{ff}^{eq}(s)$ are identical at high frequency: at $T=25\celsius$ and $T=37\celsius$, the system is in equilibrium at short time scale. Oppositely, at long time scale, the distance from equilibrium $\theta(s)$ is significantly higher at $37\celsius$ than at $25\celsius$ (figure~\ref{fig:Celsius}B). Like in the previous section, we calculated the force auto-correlation function in this regime: $\langle f_r(t+\tau) f_r(t) \rangle_t = c (\tau / \tau_0)^\gamma $. Taking $\tau_1=10~\second$, we found: $c_{37\celsius} \approx 7.6~\pico\newton^2  > c_{25\celsius} \approx 4.8~\pico\newton^2$, and $\gamma_{37\celsius} \approx 0.8 > \gamma_{25\celsius} \approx 0.5$. Both $c$ and $\gamma$ appear to increase with temperature.

\subsubsection{Discussion}
Previous microrheology measurements using magnetic microbeads have shown that the cellular medium becomes softer~\cite{vlahakis_role_2002,stroetz_validation_2001} and more dissipative~\cite{bursac_cytoskeletal_2005} as the temperature increases. Various mechanisms may explain this \cite{sunyer_temperature_2009}. In the soft glassy material mode \cite{sollich_rheology_1997,sollich_rheological_1998}, the cytoskeleton is seen as a network of polymer chains, and local rearrangements are due to binding/unbinding or unfolding of crosslinkers~\cite{hoffman_fragility_2007}. The kinetics of these reactions is thermally activated~\cite{hoffman_fragility_2007}, thus a temperature rise promotes such rearrangements, and leads to a more compliant medium. Oppositely, Sunyer \textit{et al.}~\cite{sunyer_temperature_2009}, by measuring the prestress of epithelial cells, have observed that the cell contractility increases with temperature. Indentation experiments, led in parallel with an AFM spherical tip, also show that at $13\celsius$, the cytoskeleton is softer and more dissipative than at physiological temperature. According to them, the thermal activation of the actomyosin machinery leads to an increase of the average force exerted by each myosin, or to a higher average number of active myosins per fiber, and thus to a cell rigidity increase.

In this work we measured the force auto-correlation function $\langle f_r(t+\tau) f_r(t) \rangle_t$ both at $T=37\celsius$ and $T=25\celsius$, and we point out its temperature dependance. Indeed, it behaves as a power law with time, with an exponent $\gamma$ larger at $T=37\celsius$ than at $T=25\celsius$. Following the model proposed by Robert \textit{et al.} \cite{robert_vivo_2010}, this may be directly related to the increasing number of molecular motors pulling simultaneously and in a coordinated way on the bead. This is consistent with the interpretation by Sunyer \textit{et al.} \cite{sunyer_temperature_2009}, \textit{i.e.} the number of myosins working together in stress fibers increases with temperature. 

To go beyond qualitative explanations, we discuss now the relationship between the effective temperature $T_{\text{eff}}$, the mechanical dissipation in the system, and the kinetics of biochemical reactions. In the cell, the energy is provided by fuel molecules like ATP: through their hydrolysis, these molecules converts chemical energy into mechanical energy. Biological systems are out-of-equilibrium because of this metabolic activity. As a first approximation, we assume that the rate of energy dissipation $\langle \mathcal{P} \rangle$ in the system is proportional to the rate $\nu_r$ of such reactions. Using the Arrhenius law, this rate depends on temperature as: $\nu_r\varpropto\exp(-E_0/k_B T)$ where $E_0$ represents a typical activation energy. 

On the other hand, following Harada and Sasa \cite{Harada_Sasa}, the
average power injected by the active forces, and dissipated in the
environment, $\langle \mathcal{P} \rangle$, is related to the distance
from equilibrium $\theta = \dfrac{T_{\rm eff}}{T}$, through:
\begin{equation}
\langle \mathcal{P} \rangle =  k_B T \int \dfrac{ds}{2 \pi i}  [\theta(s)-1] 
 \label{Harada-Sasa}
\end{equation}
In practice $\theta(s)-1$ is dominated by the active regime, in which
it scales as $A_2/J_0 \times s^{\beta-\alpha_2}$. Noticing that the
exponent remains almost constant as the temperature changes from
$25\celsius$ to $37\celsius$, the change in dissipated power is given
by that of $A_2/J_0$, which doubles between $25\celsius$ and
$37\celsius$. Consequently, one derives a simple expression allowing
to calculate an activation energy $E_0$:
\begin{equation}
\dfrac{E_0}{k_B}(\dfrac{1}{T_1}-\dfrac{1}{T_2}) \simeq \ln \dfrac{\langle \mathcal{P}(T_2)\rangle}{\langle \mathcal{P}(T_1) \rangle} \simeq \ln \dfrac{[A_2/J_0] (T_2)}{[A_2/J_0] (T_1)} \simeq  \ln(2)
 \label{Energy_barrier}
\end{equation}
from which one infers $E_0\approx 17 k_B T$. This value is quite consistent with typical activation energies involved in biochemical reactions: $ 8 k_B T < E_0 < 48 k_B T$, with an average of $ 24 k_B T $~\cite{Gillooly_2001}. This gives strength to our description relating mechanical dissipation to biochemical activity.

\subsection{Biological inhibitors}
In order to decipher some biological processes involved in the generation of active forces, we have targeted three active mechanisms involving the cytoskeleton dynamics and we perturbed them by pharmacologic treatments. First, the actin network itself can generate forces through dynamical remodelling of its structure. We used latrunculin A, a molecule able to bind to free G-actin monomers, in order to inhibit the actin polymerization cycle. Secondly, ATP provides the biochemical energy source necessary for actin polymerization and for myosin force production. ATP depletion was achieved by using a solution of deoxyglucose and NaN$_3$ \cite{flatman_effects_1991}. Finally, the myosins are a family of molecular motors which bind to actin filaments and pull on them. We inhibited myosin II activity by using blebbistatin~\cite{kovacs_mechanism_2004}.

\subsubsection{Results}
\paragraph*{Disruption of actin polymerization\\}

\begin{figure}[H]
  \begin{center}
  	\includegraphics[width=\columnwidth]{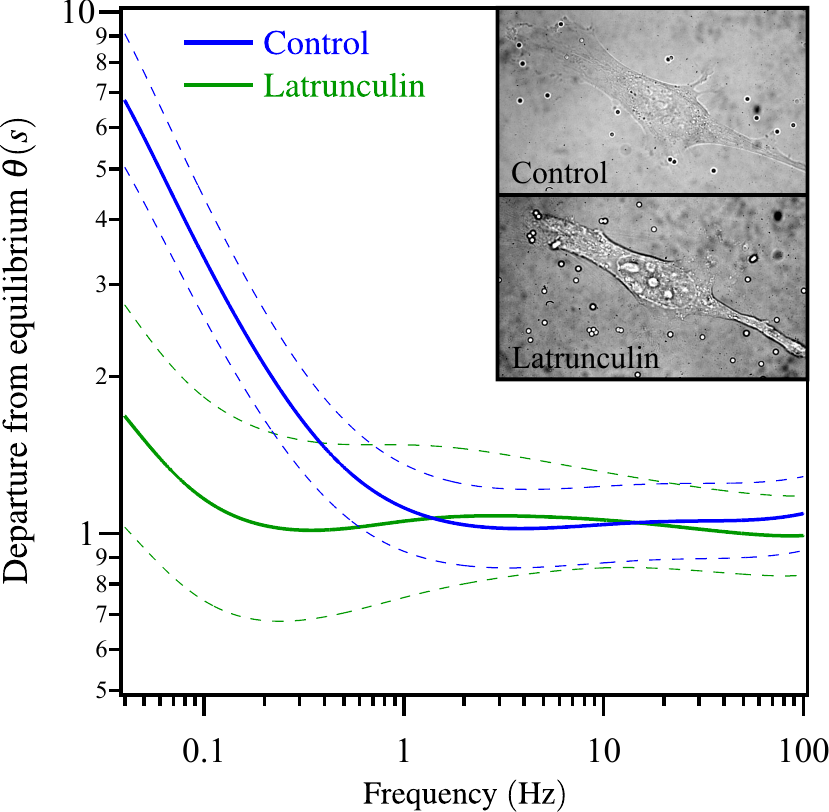}
   \caption{Effect of latrunculin A on the distance from equilibrium $\theta(s)$. The data are geometrically averaged over N = 37 realizations for control cells (blue) and over N = 10 realizations for treated cells (green). Inset: picture of a C2C12 cell before and after latrunculin treatment.\label{fig:Latrunculine}}
  \end{center}
\end{figure}

We compare the results of active and passive microrheology experiments for control cells and cells treated with latrunculin. Beads were coated with RGD 1/40. All the data presented in this section were obtained at room temperature. In the inset of fig~\ref{fig:Latrunculine} are shown two pictures of a C2C12 cell, before and after latrunculin treatment. After adding latrunculin, one observes a rounding of the cell shape, likeky associated to a decrease of the cell tension. As described in section~\ref{sec:methods}, we computed the average prefactors and exponents derived from the power law fits for the response function $J(t)$ and for the MSD $\Delta r^2(t)$. Adding latrunculin leads to an increase of the prefactor $\overline{J_0}^{Lat} = 21\pm7~\nano\meter\per\pico\newton~(N=10) > \overline{J_0}^{Ctrl} = 15\pm7~\nano\meter\per\pico\newton~(N=37)$, meaning that the cell becomes more compliant. At short time scale (subdiffusive regime), the bead MSD amplitude also increases after latrunculin addition: $\overline{A_1}^{Lat} > \overline{A_1}^{Ctrl}$ (Table~\ref{tableau_latrunculin}). Oppositely, at large time scale, the MSD amplitude $\overline{A_2}$ is slightly reduced. We also computed the force power spectrum $\hat{C}_{ff}(s)$ and the distance from equilibrium $\theta(s)$ for control cells and treated cells. Figure~\ref{fig:Latrunculine} shows that cells treated with latrunculin are close to equilibrium, at any time scale. Active (non thermal) forces are only relevant at time scales larger than $10~\second$, but their amplitude is drastically reduced as compared to control cells. 

\begin{table}[h]
\small
  \caption{Comparison of the amplitude of the response function $J(t)$ and of the MSD, for control cells and cell treated with latrunculin (\textcolor{red}{RGD 1/40})}
  \label{tableau_latrunculin}
  \begin{tabular*}{0.48\textwidth}{@{\extracolsep{\fill}}lll}
 \hline    		 
    		\small \textbf{Actin disruption RGD 1/40} & control & latrunculin \\    		
    		\hline
    $\overline{J_0}$ ($nm/pN$)&  $15 \pm 7$ &  $21 \pm 7$\\    		 
    $\overline{A_1}$ ($nm^2$) & $309\pm 46$  & $423\pm 136 $\\
    $\overline{A_2}$ ($nm^2$) & $133\pm 25$  & $97\pm 88$  \\
       
    \hline
  \end{tabular*}
\end{table} 
 
Similar experiments were performed for a different coating density: RGD 1/20. We found that latrunculin treatment increases the compliance by a factor $2.4$ ($\overline{J_0}^{Lat} = 9.9 \pm 2.6~\nano\meter\per\pico\newton > \overline{J_0}^{Ctrl} = 4.2 \pm 0.8~\nano\meter\per\pico\newton$). Similar results were obtained for the MSD amplitude ($\overline{A_1}^{Lat} = 308\pm83 ~\nano\meter^2 > \overline{A_1}^{Ctrl} = 77 \pm8 ~\nano\meter^2 $). However, no significant difference between control cells and cells treated with latrunculin was detected on $\theta(s)$ . This rather surprising result will be discussed in section \ref{Discussion: biological inhibitors}.

\paragraph*{ATP depletion\\}

\begin{figure}[h]
  \begin{center}
  	\includegraphics[width=\columnwidth]{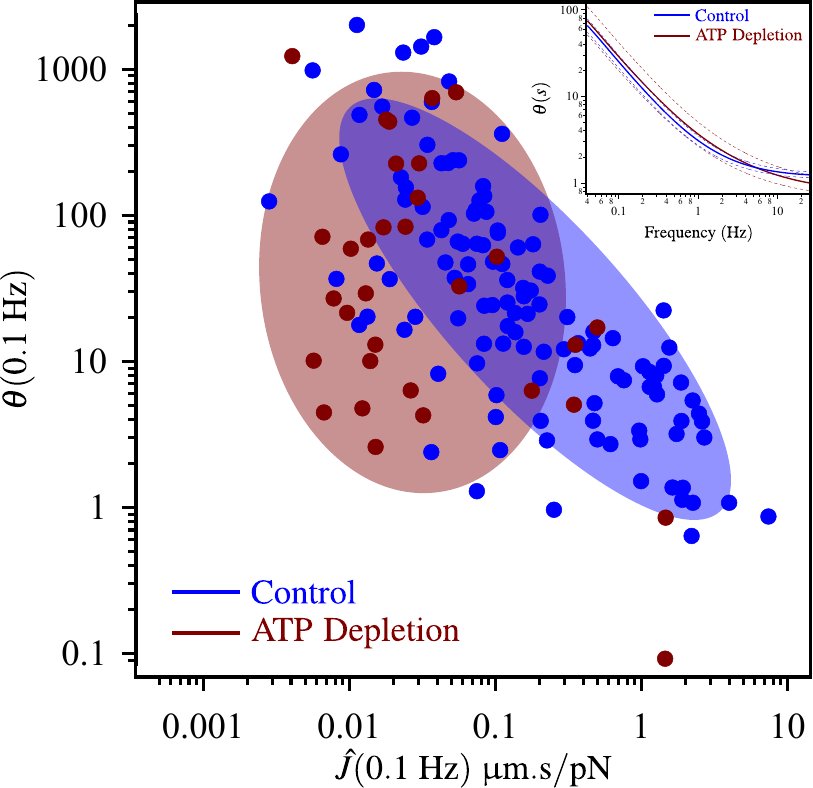}
   \caption{Scatter plot of the distance from equilibrium $\theta(s_1)$ versus response function value $\hat{J}(s_1)$, at $s_1 = 0.1~\hertz$ ($\tau_1 = 10~\second$) for control cells (blue dots) and ATP depleted cells (black dots). The ellipses are drawn according to a principal component analysis on the logarithm of $\hat{J}(s_1)$ and $\theta(s_1)$, after removing the $10\%$ most distant points from the average. Inset: mean distance from equilibrium $\hat{\theta}(s)$, geometrically averaged over $N=124$ realizations for control cells and $N=32$ realizations for ATP depleted cells.~\label{fig:ATP}}
  \end{center}
\end{figure} 

We investigated the consequences of ATP depletion on the distributions of the mechanical observables and on the distance from equilibrium. We performed this study at physiological conditions, namely at $37\celsius$. We only show here data obtained with a coating density RGD 1/20. Indeed, data at RDG 1/40 - not shown - superimpose quite well with those at RGD 1/20. Active microrheology shows that the intracellular medium is stiffer after ATP depletion (Table~\ref{tableau_ATP}). Furthermore, the exponent $\overline{\beta}$ is smaller for ATP depleted cells, indicating that their mechanical response is more solid-like and less dissipative. Concerning passive measurements, ATP depletion induces a dramatic reduction of the bead spontaneous motion: in the subdiffusive regime, for $\tau < 0.04 \second$, the MSD falls below the noise detection level ($\approx 10^{-5}~\mu m^2$); in the superdiffusive regime, ATP depletion causes a neat decrease of the MSD amplitude $\overline{A_2}$ and an increase of the MSD exponent $\overline{\alpha_2}$. 
 
\begin{table}[h]
\small
  \caption{Comparison of the active and passive microrhology parameters, for control cells and ATP depleted cells ($T = 37\celsius$; RGD 1/20)}
  \label{tableau_ATP}
  \begin{tabular*}{0.48\textwidth}{@{\extracolsep{\fill}}lll}
    \hline
     & control & ATP depletion \\
    \hline
    $\overline{J_0}$ ($nm/pN$)&  $7.9 \pm 1.2$ &  $1.3 \pm 0.3$\\   
    $\overline{\beta}$&$\approx 0.29$&$\approx 0.18$\\     
    $\overline{A_2}$ ($nm^2$) & $188\pm 19$  & $22\pm 12$  \\
    $\overline{\alpha_2}$&$\approx 1.44$&$\approx 1.53$\\   
    \hline
  \end{tabular*}
\end{table} 
 
Like in the previous sections, we combined the microrheology measurements to compute the distance from equilibrium index $\theta(s)$. In the inset of the figure~\ref{fig:ATP}, the averaged function $\theta(s)$ is plotted for ATP-depleted cells (red curve) and control cells (blue curve). Surprisingly, averaged values do not show any noticeable difference between control and ATP-depleted cells. However, one should keep in mind the natural dispersion of mechanical properties for a pool of cells in control conditions: both the compliance $J(s)$ and the out-of-equilibrium indicator $\theta(s)$ may vary over several orders of magnitude, as shown in figure~\ref{fig:J_vs_Teta_RGD}, but remain strongly correlated. Besides, ATP depletion is known not only to inhibit the cell metabolism, but also to make the cytoskeletal network stiffer, by clamping the motors onto the filaments (\textit{rigor mortis}). Thus, a meaningful comparison between the activity of control cells and of ATP-depleted cells can only be infered from the comparison of $\theta(s)$ at the same rigidity $J(s)$. Consequently, we plotted in figure~\ref{fig:ATP} the individual values of $\theta(s_1)$ vs $J(s_1)$, for a given frequency $s_1 = 0.1~\hertz$, for $N=124$ cells in control conditions (blue dots) and $N=32$ cells ATP-depleted (red dots). Like in figure~\ref{fig:J_vs_Teta_RGD}, the two ellipses are drawn according to a principal component analysis on the logarithm of $\hat{J}(s_1)$ and $\theta(s_1)$. One notices that the two clouds of points do not overlap: for a same cell compliance $\hat{J}(s)$, an ATP depleted cell will exhibit a lower distance from equilibrium $\theta(s)$ than a control cell. This will be further discussed in section \ref{Discussion: biological inhibitors}. 

\paragraph*{Myosin II inhibition \\}
\label{MyosinIIinhibition}
\begin{figure}[h]
  \begin{center}
  	\includegraphics[width=\columnwidth]{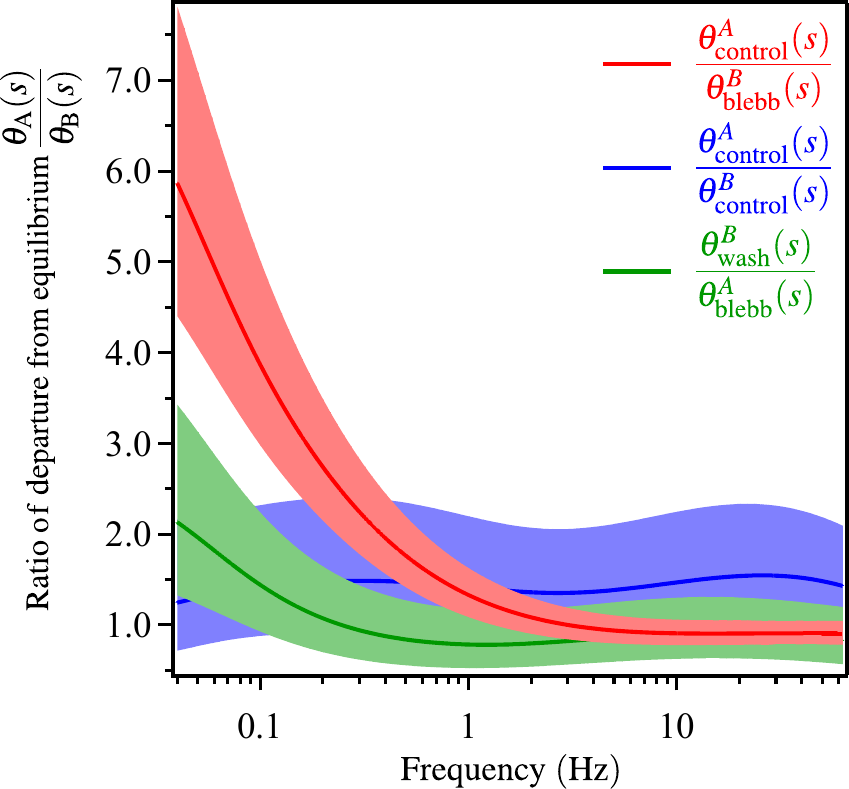}
   \caption{Effect of blebbistatin addition on the distance from equilibrium. Red curve: plot of $\theta^A_{ctrl}(s) / \theta^B_{blebb}(s)$ \textit{vs} frequency $s$, where $\theta^A_{ctrl}$ is the distance from equilibrium for control cells and $\theta^B_{blebb}$ after blebbistatin addition (average over N = 39 realizations).  Green curve: plot of the ratio $\theta^B_{wash}(s) / \theta^A_{blebb}(s)$ for cells treated with blebbistatin, and after washing drug out (N = 7). Blue curve: plot of the ratio $\theta^A_{ctrl}(s) / \theta^B_{ctrl}(s)$ for two measurements separated by 30 minutes in control conditions (N = 6).}
   ~\label{fig:Blebb}
  \end{center}
\end{figure} 

We investigate here how the generation of active forces is affected by adding blebbistatin, a well-known inhibitor of myosin II activity. A first attempt led to a negative result: the mean force power spectrum, averaged over a pool of several cells, was not significantly modified by blebbistatin addition (data not shown). Like for ATP depletion, this is due to the fact that blebbistatin affects not only the acto-myosin activity, but also the cell mechanical properties. In order to get rid of this coupling, we had to compare the data obtained on single cells, before and after drug addition. In a first serie of assays, we performed active and passive microrheology on the same bead attached to the same single cell, in control conditions (A), and 30 min after blebbistatin addition (B). In a second serie, to check the capacity of the cell to recover after myosin II reactivation, measurements were made first on a cell treated with blebbistatin (A), then the drug was washed out and a second measurement was performed 30 minutes later (B). Finally control assays were achieved without any drug addition, simply waiting for 30 minutes between the two measurements (A) and (B).

For each cell, we computed the ratio $\theta^A(s) / \theta^B(s)$ of distances from equilibrium, for the two successive measurements (A) and (B). This ratio, averaged over N cells, is plotted in figure~\ref{fig:Blebb}. The red curve corresponds to control-blebbistatin assays $\theta^A_{blebb}(s) / \theta^B_{ctrl}(s)$ (N = 39), the green one to the reverse blebbistatin-control assays $\theta^B_{wash}(s) / \theta^A_{blebb}(s)$ (N = 7), and the blue one to the control assays $\theta^A_{ctrl}(s) / \theta^B_{ctrl}(s)$ (N = 6). 

For the three series, this ratio is equal to 1 at $s\gtrsim 1Hz$, which confirms that myosin activity is never relevant in this time range ($\theta(s)=1$). For the control-blebbistatin assays, the ratio departs from 1 as $s \lesssim 1Hz$, and reaches $\approx 6$ at $s = 0.04 Hz$: this indicates a neat decrease of the amplitude of active forces when the myosin II activity is inhibited. Oppositely, in the blebbistatin-control sequence, the ratio $\theta^B_{wash}(s) / \theta^A_{blebb}(s)$ slightly increases by a factor $\approx 2$, at low frequencies, indicating a partial recovery of myosin II activity after blebbistatin washing out. However the small number of measurements in this sequence (N=7) does not allow a definitive conclusion on this recovery. Finally, in the control-control sequence, the ratio $\theta^A_{ctrl}(s) / \theta^B_{ctrl}(s)$ remains almost constant, equal to 1, in the complete frequency range. This confirms that, in the absence of any external perturbation, the biological activity index is not modified during the 30 min time lapse between (A) and (B). This conforts the validity of the procedure.

\subsubsection{Discussion}
\label{Discussion: biological inhibitors}
We have tested in this work three different pathways to perturb the cell normal activity, and we measured the consequences on the distribution of active forces generated in the cortex. First latrunculin A was used to inhibit actin polymerization, and thus to disrupt actin filaments. After latrunculin addition, the cell medium is more compliant, consistently with the literature~\cite{laudadio_rat_2005,yamada_2000,robert_vivo_2010}. Partial depolymerization of the actin network around the bead softens the bead environment, and consequently the bead spontaneous motion generated by thermal forces is amplified (see section \ref{sec:MSDscaling}, Eq.~\eqref{eq:thermal}). However, adding latrunculin has no systematical effect on the bead directed motion: for instance, for a ligand density RGD 1/40, the superdiffusive regime is suppressed and the force power spectrum is similar to the equilibrium one. On the other hand, for a ligand density RGD 1/20, the bead is sensitive to active forces and its motion remains partially directed. We conclude that a higher degree of actin polymerization in the cortex contributes to a more efficient transmission of active forces to the bead. This transmission depends on two factors: the connectivity of the actin network (which is impaired by depolymerizing drugs), and the binding between the bead and the actin network (which depends on the ligand density, see Eq.~\eqref{eq:active}). These two factors might cooperate to increase the transport efficiency. 

ATP depletion is well-known to freeze the cell metabolism, but also to increase the cell stiffness. Indeed, in the absence of ATP, the myosins heads remain bound to the actin filaments and crosslink them, which makes the network more rigid. ATP depletion is also reported to induce an increasing of actin polymerisation on epithelial~\cite{atkinson_mechanism_2004} and endothelial~\cite{suurna_cofilin_2006} cells. According to~\cite{suurna_cofilin_2006}, ATP depletion may result in destruction of actin stress fibers and accumulation of F-actin aggregates. 

Actually, our results for the distance from equilibrium $\theta (s)$ show no significant differences between averaged control cells and averaged ATP depleted cells (figure~\ref{fig:ATP} inset). Thus, to interpret correctly the data, it is mandatory to take into account the rigidity increase associated to ATP depletion. Indeed, figure \ref{fig:J_vs_Teta_RGD} shows that the distance from equilibium is correlated to the cortex rigidity in the bead vicinity: if the medium to which the bead is connected appears stiffer (due either to a higher ligand density or to a local higher rigidity of the actin network) the resultant forces transmitted to the bead should be larger, and consequently the measured distance from equilibrium is higher. This appears in figure~\ref{fig:ATP}, where the points corresponding respectively to control cells and to ATP depleted cells divide in two separate clusters. We conclude that, although the expected effect may be partially hidden by cell stiffening, ATP depletion actually reduces the amplitude of active forces generated in the cell cortex. 

However, in our experiment, active force generation is not totally suppressed by ATP depletion, since the force spectrum after treatment does not coincide with the thermodynamical equilibrium one. Two factors might explain this: either the ATP concentration is not lowered enough to eliminate all ATP-dependant mechanical activity, or other mechanisms of force generation, which are not ATP dependant (like for instance GTP-activated microtubule polymerization) also contribute to the bead active motion. 

Finally, we could confirm that myosin II activity is one of the main force generation mechanism in the actin cortex: indeed, the inhibition of myosin II by blebbistatin reduces the auto-correlation function of forces exerted on the probes, by a factor up to 6, in a reversible manner (see figure~\ref{fig:Blebb}). However, as mentioned above, myosins may not be the exclusive source of active forces in the actin cortex, but their contribution appear to be dominant. This enforces the consistency of our description of the force generation processes in the actin cortex, where myosin II acts as the main motor, ATP is the main source of energy and actin filaments are the tracks for the force transmission. The force amplitude derived in section~\ref{Force power spectrum} also indicates that the bead motion is driven by only a few individual motors simultaneously pulling on it.

\section{Final remarks and Conclusion}

This work contributes to make quantitative the description of the active forces generated and transmitted in the cortex of living cells, through the conversion of biochemical into mechanical energy. On the experimental side, the amplitude and frequency dependence of the force auto-correlation function is quantitatively measured by combining active and passive rheological experiments on the same probe, namely a micron-sized bead specifically bound to the actin cortex through the membrane adhesive receptors. For cells in control conditions, the stochastic force distribution is dominated by thermal fluctuations at short time scale, where the fluctuation-dissipation theorem is verified. Oppositely, at long time scale, the force distribution is dominated by active non-thermal forces. In this case the fluctuation-dissipation theorem no longer applies and the deviations from thermal equilibrium are quantified. A theoretical approach is developed, which takes into account the binding of the bead to the cortex and the visco-elastic properties of the medium, and leads to an accurate description of the different regimes of the bead motion. Actin filaments disruption, ATP depletion and myosin II inhibition allows us to identify the acto-myosin molecular motor as the main source of the active forces, in the nanonewton range. Both the influence of the temperature and of the density of bonds on the bead are carefully investigated. All the data are made consistent with each other through a global description of the active mechanisms responsible for the probe motion. In summary, in this work we give a detailed, quantitative, and time-resolved picture of the distribution of stochastic forces in the actin cortex of the cell, and we propose a consistent interpretation of the different active mechanisms responsible for their generation and transmission along the cytoskeleton.



\balance


\bibliography{rsc} 
\bibliographystyle{rsc} 

\end{document}